\documentclass[fleqn,usenatbib]{mnras}

\usepackage[T1]{fontenc}

\DeclareRobustCommand{\VAN}[3]{#2}
\let\VANthebibliography\thebibliography
\def\thebibliography{\DeclareRobustCommand{\VAN}[3]{##3}\VANthebibliography}

\usepackage{graphicx}	
\usepackage{amsmath}	
\usepackage{amssymb}	
\usepackage{soul}
\usepackage{gensymb}
\usepackage{hyperref}
\usepackage{subcaption}

\def\mass{$M_{\rm{\odot}}$}

\newcommand{\nickel}{$^{56}$Ni}
\newcommand{\iron}{$^{52}$Fe}
\newcommand{\chromium}{$^{48}$Cr}
\newcommand{\titanium}{$^{44}$Ti}
\newcommand{\calcium}{$^{40}$Ca}
\newcommand{\argon}{$^{36}$Ar}
\newcommand{\sulphur}{$^{32}$S}
\newcommand{\silicon}{$^{28}$Si}
\newcommand{\magnesium}{$^{24}$Mg}
\newcommand{\neon}{$^{20}$Ne}
\newcommand{\oxygen}{$^{16}$O}
\newcommand{\nitrogen}{$^{14}$N}
\newcommand{\carbon}{$^{12}$C}

\usepackage{newtxtext,newtxmath}

\title[The diversity of double detonation explosions]{Exploring the diversity of double detonation explosions for type Ia supernovae: Effects of the post-explosion helium shell composition }

\author[M. R. Magee et al.]{
M. R. Magee$^{1,2}$\thanks{E-mail: mrmagee.astro@gmail.com},
K. Maguire$^{1}$,
R. Kotak$^{3}$,
S. A. Sim$^{2}$
\\
$^{1}$School of Physics, Trinity College Dublin, The University of Dublin, Dublin 2, Ireland \\ 
$^{2}$Astrophysics Research Centre, School of Mathematics and Physics, Queen's University Belfast, Belfast, BT7 1NN, UK\\
$^{3}$Tuorla Observatory, Department of Physics and Astronomy, FI-20014 University of Turku, Finland
}

\date{Accepted 2021 January 13. Received 2021 January 12; in original form 2020 August 24.}

\pubyear{2021}

\begin{document}
\label{firstpage}
\pagerange{\pageref{firstpage}--\pageref{lastpage}}
\maketitle

\begin{abstract}
The detonation of a helium shell on top of a carbon-oxygen white dwarf has been argued as a potential explosion mechanism for type Ia supernovae (SNe~Ia). The ash produced during helium shell burning can lead to light curves and spectra that are inconsistent with normal SNe~Ia, but may be viable for some objects showing a light curve bump within the days following explosion. We present a series of radiative transfer models designed to mimic predictions from double detonation explosion models. We consider a range of core and shell masses, and systematically explore multiple post-explosion compositions for the helium shell. We find that a variety of luminosities and timescales for early light curve bumps result from those models with shells containing \nickel{}, \iron{}, or \chromium{}. Comparing our models to SNe~Ia with light curve bumps, we find that these models can reproduce the shapes of almost all of the bumps observed, but only those objects with red colours around maximum light ($B-V \gtrsim 1$) are well matched throughout their evolution. Consistent with previous works, we also show that those models in which the shell does not contain iron-group elements provide good agreement with normal SNe~Ia of different luminosities from shortly after explosion up to maximum light. While our models do not amount to positive  evidence in favour of the double detonation scenario, we show that provided the helium shell ash does not contain iron-group elements, it may be  viable for a wide range of normal SNe~Ia.

\end{abstract}




\section{Introduction}
\label{sect:intro}

One of the most debated aspects of research on type Ia supernovae (SNe~Ia) is whether multiple progenitor systems are needed to explain the entire population (see \citealt{livio--18, wang--18, jha--19, soker--19} for recent reviews of SNe~Ia). Despite significant work throughout the years, the question remains whether SNe~Ia primarily result from Chandrasekhar or sub-Chandrasekhar mass white dwarfs.

\par

To trigger the detonation of a sub-Chandrasekhar mass white dwarf, early models invoked scenarios in which a massive helium shell ($\lesssim$0.2~\mass{}) accumulates on the surface of the white dwarf (e.g. \citealt{livne--90, livne--91, woosley--94}). As the mass of the helium shell increases through accretion, the density and temperature at the base of the shell also increase. Eventually convective nuclear burning may develop and potentially transition to a detonation. Following ignition of the shell, a secondary detonation may be triggered in the core. This secondary detonation can be triggered in multiple ways (converging shock, edge-lit, or scissors mechanism; \citealt{livne--90, livne--91, moll--13, gronow--20}), however the end result is the same -- complete disruption of the white dwarf. This is the so-called double detonation scenario. Within these models, most studies find that burning in the helium shell proceeds mostly to nuclear statistical equilibrium (NSE) -- producing a large amount of \nickel{} and other iron-group elements (IGEs). Such a large mass of IGEs in the outer ejecta leads to significant line blanketing that generally does not agree with observations of SNe Ia \citep{hoeflich--96a, nugent--97}.

\par

Given the adverse impact of the helium shell ash on the light curves and spectra, there has been significant interest in minimising its effects. Neglecting any helium shell altogether, models invoking pure detonations of isolated, bare sub-Chandrasekhar mass white dwarfs have been shown to broadly reproduce the light curves and spectra of normal SNe Ia \citep{sim--10, shen--18, goldstein--18}. Such white dwarfs however, will not spontaneously detonate and therefore these explosions do not occur naturally.  Alternatively, models with thin helium shells may also be a viable pathway to explain normal SNe~Ia. \citet{bildsten--07} showed that ignition within the helium shell can be achieved for much lower masses of $\sim$0.02~\mass{}, but they did not not consider the possibility of core ignition following the initial helium shell detonation. Subsequent core ignition was shown to be robustly achieved by \citet{fink--07}, \citet{fink--10}, and \citet{shen--14} for high-mass white dwarfs ($\gtrsim$0.8~\mass{}). In spite of these lower shell masses, models presented by \citet{kromer--10} and \citet{gronow--20} remain inconsistent with the observed light curves and spectra of normal SNe Ia. 
Recently, \citet{polin--19} presented a suite of double detonation models covering a range of core and shell masses (from 0.6 -- 1.2~\mass{} and 0.01 -- 0.1~\mass{}, respectively) and find that some models with thin helium shells do produce spectra that resemble normal SNe~Ia.

\par

In addition to producing strong line blanketing, the presence of IGEs in the helium shell ash has an important consequence for the light curves predicted by double detonation explosions. \citet{noebauer-17} and \citet{jiang--2017} have shown that the production of short-lived radioactive isotopes (\nickel{, }\iron{}, and \chromium{}) in the shell results in a distinct bump in the early light curve (within approximately three days of explosion). Studies of samples of SNe~Ia (e.g. \citealt{bianco--11, olling--15, papadogiannakis--2019, miller--20a}) have shown that the evidence for clear bumps is relatively rare, but a few candidate objects have been proposed (e.g. \citealt{jiang--2017, hosseinzadeh--17, li--19, de--19, miller--20}). 

\par

Qualitatively similar bumps in the early light curves of SNe Ia are also suggested to be produced via different mechanisms, such as the presence of a \nickel{} excess in the outer ejecta \citep{magee--20b}, interaction with a companion star \citep{kasen--10}, or interaction with circumstellar material (CSM; \citealt{piro-16}). An excess of \nickel{} in the outer ejecta may result from plumes of burned ash rising to the surface of the white dwarf during explosion. As the \nickel{} decays to $^{56}$Co, the radiation produced is able to quickly escape from the ejecta surface and results in a light curve bump. The luminosity and duration of the bump depends on both the mass and distribution of \nickel{}. In the interaction scenarios, a light curve bump may be produced due to cooling of the shocked ejecta following the interaction. For companion interaction, the bump is affected by the nature of the companion, with more evolved stars producing stronger interaction signatures. In both cases, the mass and extent of the interacting material will also determine the luminosity and duration of the bump. 

\par

\citet{maeda--18} specifically investigate the different early light curve signatures predicted by the double-detonation scenario and interaction. The models presented by \citet{maeda--18} show significant overlap between these two scenarios, in terms of the duration and luminosity of the bump, but the double detonation in general produces somewhat redder colours. \citet{maeda--18} show that this is at least partially due to the specific IGEs present in the shell. 

\par

Aside from the mass of the helium shell, it has also been suggested that its composition can play an important role during nuclear burning, and can dramatically affect the post-explosion observable properties. \citet{kromer--10} presented a model in which the helium shell was polluted by carbon (34\% by mass) and found that burning within the shell did not proceed to NSE, but instead stalled earlier in the $\alpha$-chain. In this case, the lack of IGE in the shell produced light curves and spectra that are generally consistent with normal SNe~Ia. In addition, \citet{townsley--19} recently showed that the inclusion of other isotopes, besides carbon, can also dramatically affect the post-explosion composition of the shell and produce observables comparable to normal SNe Ia. Therefore, there is considerable scope for variation in the burning products produced in the shell.

\par

In this work, we present radiative transfer simulations exploring a range of ejecta models that are designed to parameterise and broadly mimic predictions from double-detonation explosion models. We perform the first large-scale exploration of various compositions for the helium shell following explosion, and determine the range of models that do and do not reproduce observations of SNe~Ia. Although different helium shell compositions in parameterised models were previously studied by \citet{maeda--18}, here we explore a wider range of compositions in the helium shell, as well as multiple shell masses for a given core mass. In Sect.~\ref{sect:rt_modelling}, we discuss the radiative transfer code used in this work, TURTLS \citep{magee--18}. Sect.~\ref{sect:construct_models} presents our approach to constructing parameterised double detonation models. In Sect.~\ref{sect:shell_composition}, we discuss the impact of the helium shell composition on the model observables, while in Sect.~\ref{sect:burned_mass} we show the impact of the mass of burned material above the core. The rise times and early light curve bumps of our models are discussed in Sect.~\ref{sect:rises}. In Sect.~\ref{sect:comparisons_ni}, we compare to existing models with varying \nickel{} distributions. Comparisons to observations of normal SNe~Ia are presented in Sect.~\ref{sect:comparisons_normal}, while in Sect.~\ref{sect:comparisons_bump} we compare to SNe~Ia showing a bump in the early light curve. For all spectral comparisons, spectra have been corrected for Milky Way and host extinction, where appropriate, and were obtained from WISeREP \citep{wiserep}. Finally, we present our conclusions in Sect.~\ref{sect:conclusions}.

%

\section{Radiative transfer modelling}
\label{sect:rt_modelling}

\begin{figure}
\centering
\includegraphics[width=\columnwidth]{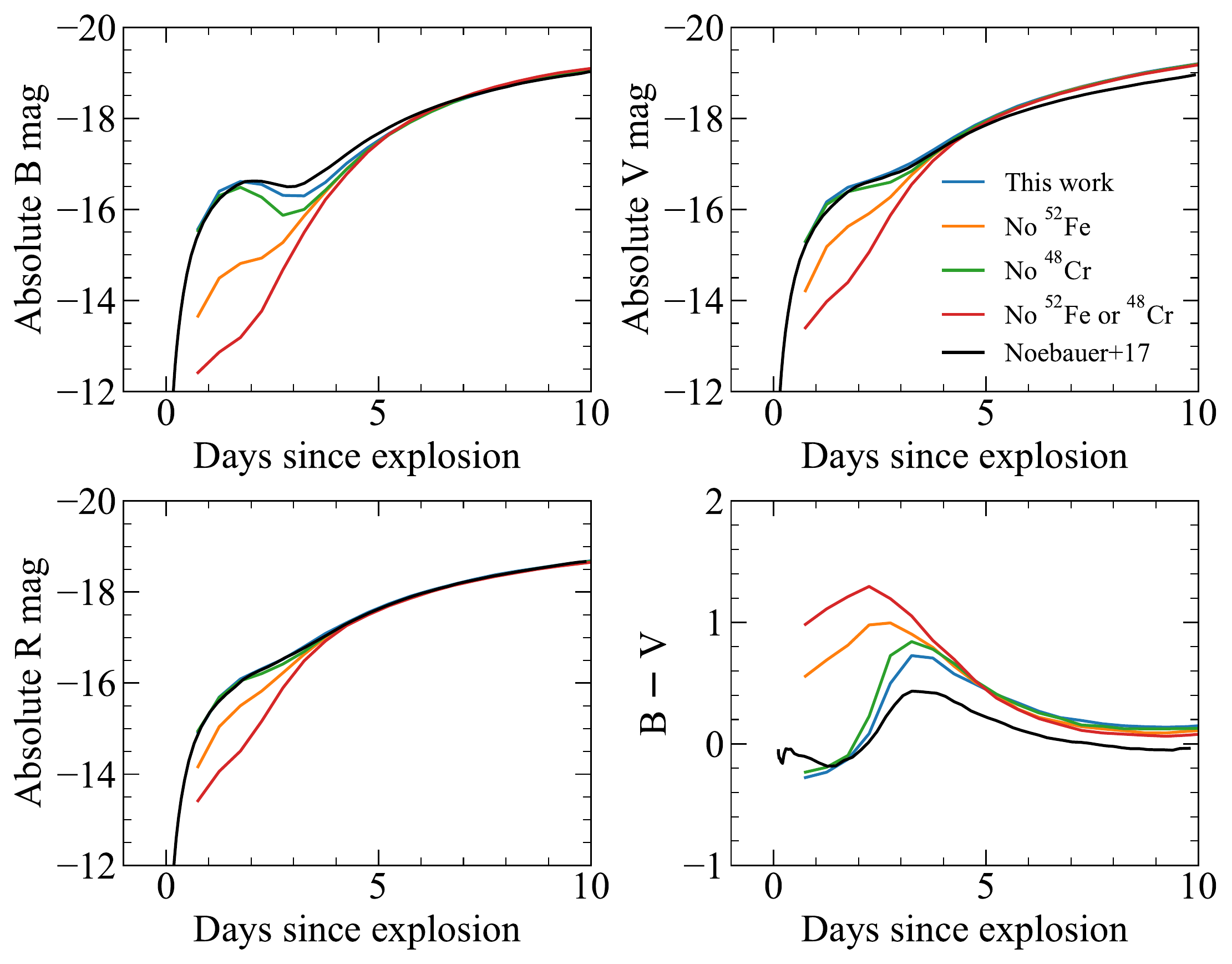}
\caption{Comparison between the 
sub-Chandrasekhar mass double detonation 
model calculated by \citet{noebauer-17} 
using STELLA (black) and our calculation 
using TURTLS (blue).} 
\label{fig:noebauer_comp}
\centering
\end{figure}

We use the one dimensional radiative transfer code TURTLS \citep{magee--18} to perform our simulations. All model light curves and spectra presented in this work are freely available on GitHub\footnote{\href{https://github.com/MarkMageeAstro/TURTLS-Light-curves}{https://github.com/MarkMageeAstro/TURTLS-Light-curves}}. TURTLS is described in detail by \citet{magee--18}. Here we provide a brief overview of the code and outline changes implemented for this study. 

\par

TURTLS is a Monte Carlo radiative transfer code following the methods of \citet{lucy-05} (see \citealt{noebauer--19}, and references therein, for a review of Monte Carlo radiative transfer methods). TURTLS is designed for modelling the early time evolution of thermonuclear supernovae. For each simulation, the density and composition of the model ejecta is defined in a series of discrete cells. Monte Carlo packets representing bundles of photons are injected into the model region, tracing the decay of radioactive isotopes. We have updated TURTLS to account for energy generated by the $^{52}$Fe $\rightarrow$ $^{52}$Mn $\protect\rightarrow$ $^{52}$Cr and $^{48}$Cr $\protect\rightarrow$ $^{48}$V $\rightarrow$ $^{48}$Ti decay chains, which can contribute significantly to the luminosity and overall evolution of the model in the double detonation scenario \citep{noebauer-17}. Isotope lifetimes and decay energies are taken from \citet{dessart--2014b}.

\par

For all simulations presented in this work, we use a start time of 0.5\,d after explosion. In the appendix in Sect.~\ref{sect:converge}, we show the results of some of our convergence tests with earlier start times. These tests demonstrate that, despite the short half-lives of many of the included isotopes, a start time of 0.5\,d after explosion does not significantly alter the synthetic observables and does not impact our conclusions. Once packets are injected into the model region, their propagation is followed until either they escape or the simulation ends. Due to the assumption of local thermodynamic equilibrium (LTE) within TURTLS, simulations are stopped at 30 days after explosion. We also note that as a consequence of this assumption, our models do not predict the presence of helium features, despite the potential for a large amount of unburned helium. Previous studies have shown that a non-LTE treatment of helium is required to produce spectral features for the conditions typical of SN ejecta \citep{hachinger--12, dessart--15, boyle--17}. At the start of each simulation, packets are injected as $\gamma$-packets (representing $\gamma$-ray photons) and treated with a grey opacity of 0.03~cm$^{2}$~g$^{-1}$. Following an interaction with the model ejecta, $\gamma$-packets are converted to optical radiation packets ($r$-packets). For these packets, we use TARDIS \citep{tardis, tardis_v2} to calculate the non-grey expansion opacities and electron-scattering opacities within each cell during the current time step. During each time step, we extract a `virtual' spectrum using the so-called event-based technique (e.g \citealt{long--knigge--02, sim--10-agn, tardis, bulla--15, magee--20b}). Light curves are calculated via the convolution of synthetic virtual spectra with the desired set of filter functions at each time step. 

\par

In Fig.~\ref{fig:noebauer_comp}, we show a comparison of our model light curves including the new decay chains to those calculated by \citet{noebauer-17}, using the radiative transfer code, STELLA \citep{stella--98, stella--06}, for the same model structure. This model involves the detonation of a 0.055~\mass{} helium shell on a 1.025~\mass{} carbon-oxygen white dwarf. The resulting explosion leads to the production of 0.55~\mass{} of \nickel{} in the white dwarf core. The helium shell ash following explosion is dominated by IGEs, which includes  $\sim$0.002~\mass{} of \nickel{}, 0.006~\mass{} of \iron{}, and 0.004~\mass{} of \chromium{}. Figure~\ref{fig:noebauer_comp} verifies that with our implementation of the additional decay chains, TURTLS can broadly match the light curves of \citet{noebauer-17}. The early light curve bump observed in our models is somewhat less pronounced than in the \citet{noebauer-17} model, which is likely a result of differences in the treatment of opacities, for example. 

\par

We also show light curves in Fig.~\ref{fig:noebauer_comp} calculated including either the $^{52}$Fe $\rightarrow$ $^{52}$Mn $\protect\rightarrow$ $^{52}$Cr chain, the $^{48}$Cr $\rightarrow$ $^{48}$V $\protect\rightarrow$ $^{48}$Ti chain, or neither, as a further demonstration of their contribution to the early luminosity. We note that in all cases, the $^{56}$Ni $\protect\rightarrow$ $^{56}$Co $\protect\rightarrow$ $^{56}$Fe decay chain is included. Including these additional chains produces a $\sim$4~mag. increase in the brightness by approximately two days after explosion. Figure~\ref{fig:noebauer_comp} shows that despite the short half-lives of both the parent and daughter isotopes ($t_{1/2}$ = 0.345\,d and 0.015d, respectively), the $^{52}$Fe $\protect\rightarrow$ $^{52}$Mn $\protect\rightarrow$ $^{52}$Cr chain contributes significantly to the early luminosity within the first few days of explosion. For this model, the early bump reaches a peak $B$-band magnitude of $-$16.7~mag. approximately 1.8\,d after explosion. At this time, the instantaneous energy deposition rate from material in the shell reaches $\sim1.3 \times 10^{42}$~erg~s$^{-1}$ and dominates the luminosity output of the model, which is consistent with expectations from Arnett's law \citep{arnett--law}.

%

\section{Constructing the double detonation model set}
\label{sect:construct_models}

In the following section, we discuss our approach to creating a parameterised description of the ejecta in double detonation explosions. Our strategy is based on capturing and exploring the variation present across a range of published models in a systematic way. Each of our models is controlled by the following parameters: the mass of the carbon-oxygen core, the mass of the helium shell, the fraction of the helium shell burned during the explosion, and the dominant $\alpha$-chain product produced in the shell burning. The range of input parameters used is shown in Table~\ref{tab:model_params}. The name of each model is also derived based on these parameters, for example WD1.00\_He0.04\_BF0.50\_DP56Ni refers to a model with a core mass of 1.0~\mass{}, a helium shell mass of 0.04~\mass{}, of which 50\% is burned, and the dominant product produced in the shell is \nickel{}. The range of parameters explored was chosen to broadly cover and bracket the values predicted by various explosion models, but we stress they are not exact reproductions of existing models.

\begin{table}
\centering
\caption{Ejecta model parameters}\tabularnewline
\label{tab:model_params}\tabularnewline
\resizebox{\columnwidth}{!}{
\begin{tabular}{cccc}
\hline
\hline
\tabularnewline[-0.25cm]
Core mass   &	Helium shell & Fraction of     & Dominant burning  \tabularnewline
            &   mass       & shell burned     & product in shell   \tabularnewline
\mass{}     & \mass{}	     &                  & 	              \tabularnewline
\hline
\hline
\tabularnewline[-0.2cm]
0.90	    &	0.01, 0.04, 0.07, 0.10                    & 0.20, 0.50, 0.80 		    & 	\sulphur{} -- \nickel{}	\tabularnewline 
1.00	    &	0.01, 0.04, 0.07, 0.10  & 0.20, 0.50, 0.80 		    & 	\sulphur{} -- \nickel{}	\tabularnewline 
1.10	    &	0.01, 0.04, 0.07, 0.10  & 0.20, 0.50, 0.80 		    & 	\sulphur{} -- \nickel{}	\tabularnewline 
1.20	    &	0.01, 0.04, 0.07, 0.10  & 0.20, 0.50, 0.80 		    & 	\sulphur{} -- \nickel{}	\tabularnewline
\hline
\hline
\end{tabular}
}
\end{table}

\begin{figure}
\centering
\includegraphics[width=\columnwidth]{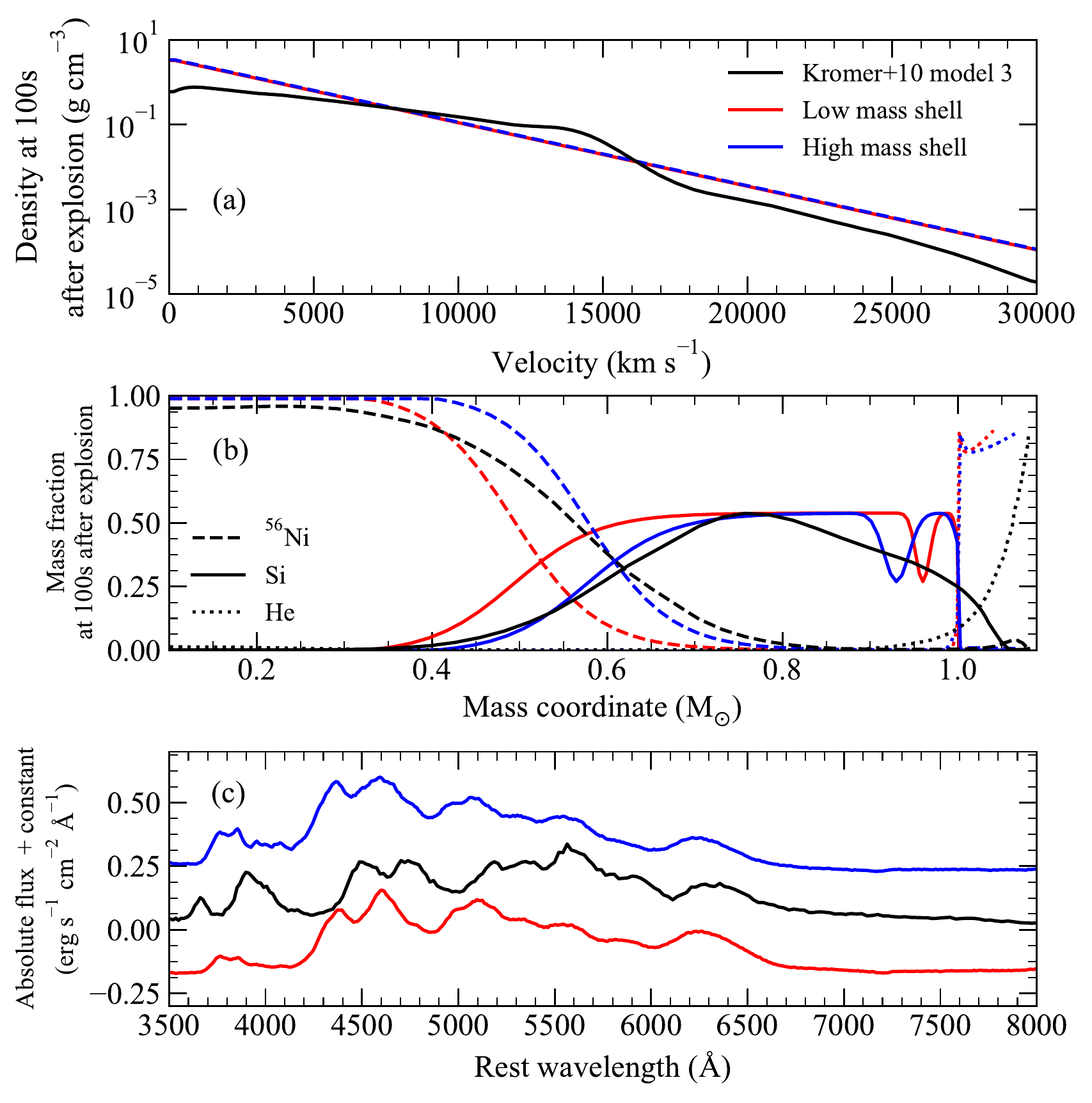}
\caption{Comparison of properties for \citet{kromer--10} model 3 (1.025~\mass{} core, 0.055~\mass{} helium shell) and our models with a 1.0~\mass{} core and 0.04~\mass{} (red) and 0.07~\mass{} (blue) helium shell. {\it Panel a:} Comparison between model density profiles. {\it Panel b:} Comparison between model compositions. {\it Panel c:} Maximum light spectra for all models (see Sect.~\ref{sect:construct_models} for further details). Spectra are offset vertically for clarity.}
\label{fig:compositions}
\centering
\end{figure}

For each model, we require a density profile for the ejecta. The density profiles presented in \citet{magee--18} and \citet{magee--20} were designed to broadly mimic those from a variety of explosion scenarios. In particular, the exponential density profile with a kinetic energy of 1.4$\times$10$^{51}$~erg from \citet{magee--20} bears a striking similarity to the models of \citet{kromer--10} and \citet{polin--19}, although the density in the outer ejecta is slightly higher. We therefore take this model as our nominal profile shape and simply scale the density to the appropriate ejecta mass, which is given by the sum of the core and helium shell masses. A demonstrative comparison between model 3 of \citet{kromer--10} and two of our models is shown Fig.~\ref{fig:compositions}(a). We note that the core mass of model 3 (1.025~\mass{}) is slightly higher than these models (1.0~\mass{}), and we show two shell masses (0.04 and 0.07~\mass{}) to bracket the 0.055~\mass{} shell of model 3.

\par

\subsection{Composition of the core}
\label{sect:construct_core_comp}
\begin{figure}
\centering
\includegraphics[width=\columnwidth]{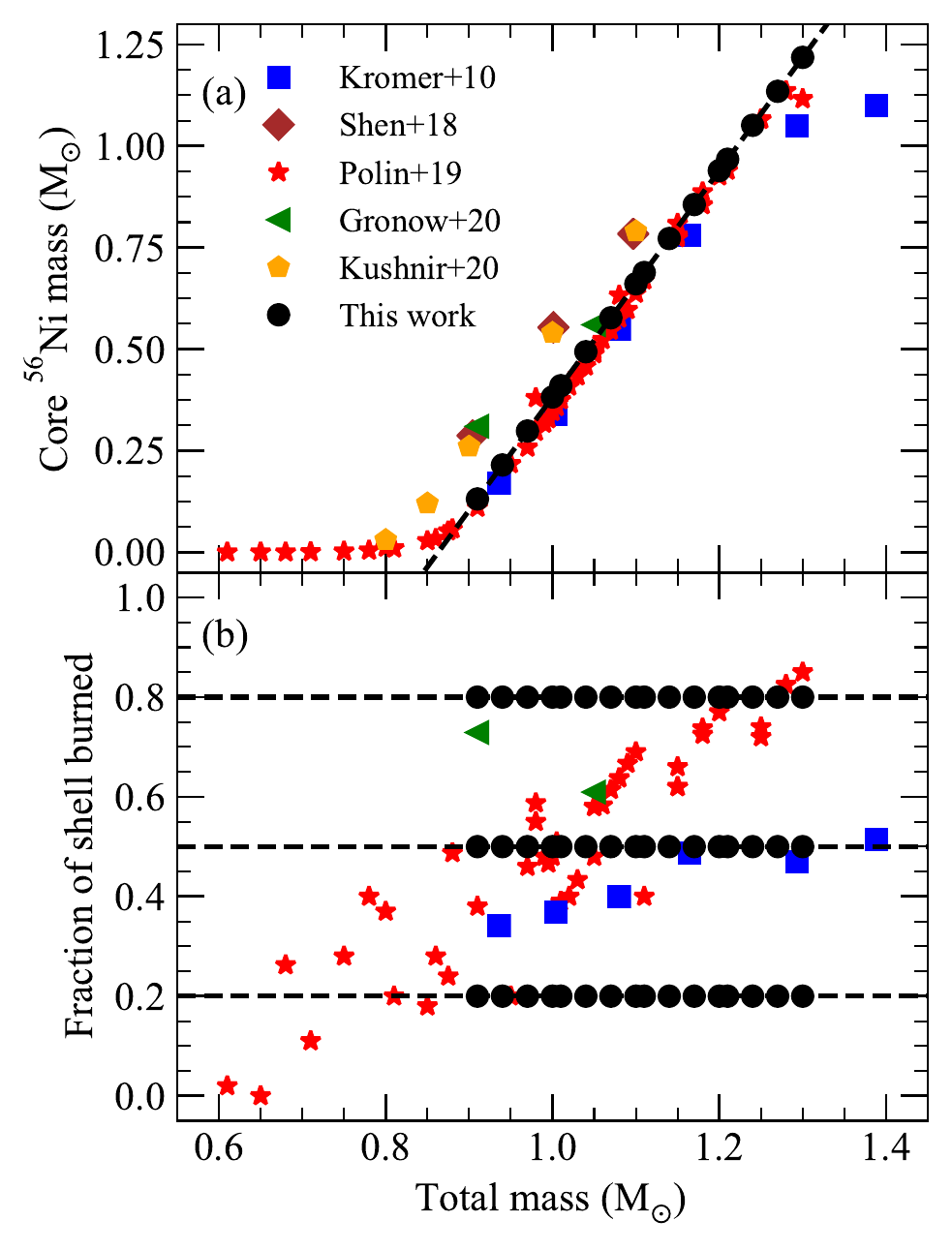}
\caption{{\it Panel a:} \nickel{} mass produced in the carbon-oxygen core as a function of total mass of the white dwarf (sum of core and shell mass). Literature values are taken from their respective papers \citep{kromer--10, shen--18, polin--19, gronow--20, kushnir--20}. For total masses between 0.90 -- 1.30~\mass{}, we show a linear fit to the \citet{polin--19} models, which is used to determine the core \nickel{} mass of our models. {\it Panel b:} Fraction of the helium shell that is burned (i.e. converted to elements heavier than helium following the explosion) as a function of total mass. Dashed horizontal lines show fractions of 0.2, 0.5, and 0.8. Black points show the specific models calculated in this work, which broadly extend the range predicted from a variety of explosion models.}
\label{fig:ni_masses}
\centering
\end{figure}

Previous studies of double detonation explosions have shown that the amount of \nickel{} produced in the carbon-oxygen core during the explosion is directly related to the total mass of the white dwarf. In Fig.~\ref{fig:ni_masses}(a) we show the core \nickel{} mass produced as a function of total mass for a sample of models from the literature \citep{kromer--10, shen--18, polin--19, gronow--20, kushnir--20}. As shown in Fig.~\ref{fig:ni_masses}(a), there is disagreement between studies over the total amount of \nickel{} produced. For example, the \citet{polin--19} models predict a \nickel{} mass of $\sim$0.4~\mass{} for a total white dwarf mass of $\sim$1.0~\mass{} whereas \citet{kushnir--20} predict $\sim$0.55~\mass{}. Models presented by \citet{kromer--10}, \citet{polin--19}, and \citet{gronow--20} focus on helium-shell detonations, while those of \citet{shen--18} and \citet{kushnir--20} are instead detonations of bare, sub-Chandrasekhar mass white dwarfs. For this reason, we we use the former set of models as reference points throughout this study, allowing us to consistently select parameters for our model helium shells and cores masses. Between $\sim$0.9 -- 1.3~\mass{} there is an approximately linear relation and broad agreement between these different model sets. As the \citet{polin--19} sample covers a large range of total masses and different ignition conditions, we use a linear fit to this model set to determine the core \nickel{} mass of our models. The core \nickel{} mass is therefore given by:
\begin{equation}
\label{eqn:ni-mass}
    M(^{56}{\rm Ni}) = 2.8 \times (M_{\rm{core}} + M_{\rm{shell}} ) - 2.4,
\end{equation}
where $M_{\rm{core}}$ is the mass of the carbon-oxygen core and $M_{\rm{shell}}$ is the mass of the helium shell. All variables are in units of \mass{}. This fit is shown as a dashed line in Fig.~\ref{fig:ni_masses}(a).

\par

In Appendix.~\ref{sect:core_ni}, we present additional models exploring \nickel{} masses based on the \citet{shen--18} and \citet{kushnir--20} models. Given the uncertainty in the amount of \nickel{} produced, the total white dwarf mass should not be taken as a prediction from our models. Throughout this work, we give the values of core and shell masses simply as reference to identify each model. Instead, we consider the total luminosity (i.e. the \nickel{} mass) to be a robust prediction and expect that there may be a range of white dwarf properties that produce such a \nickel{} mass.

\par

For the distribution of \nickel{} within the core, we follow the functional form used by \citet{magee--18}. The $^{56}$Ni mass fraction at mass coordinate $m$ is given by:
\begin{equation}
\label{eqn:ni_dist}
^{56}{\rm Ni}\left(m\right) = \frac{1}{\exp\left(s\left[m - M_{\rm{Ni}}\right]/M_{\rm{\odot}}\right) + 1 },
\end{equation}
where $M_{\rm{Ni}}$ is the total \nickel{} mass in units of \mass{}. The scaling parameter, $s$, is used to control how quickly the ejecta transitions from a \nickel{}-rich to -poor composition. The models with $s = 21$ presented by \citet{magee--20} produce a \nickel{} distribution qualitatively similar to those of \citet{kromer--10} (Fig.~\ref{fig:compositions}(b)), therefore we fix $s = 21$ for all models in this work. By adopting a similar method to \citet{magee--20} and \citet{magee--20b}, we also allow for a direct comparison to the models presented in both studies. Immediately below the base of the helium shell, we place a small amount ($\sim$10$^{-3}$ -- 10$^{-1}$~\mass{}) of unburned carbon and oxygen assuming a Gaussian distribution (width $\sim$0.001 -- 0.01~\mass{}). The mass and distribution of this unburned material is comparable to that predicted by explosion models (e.g. \citealt{kromer--10, polin--19}), although in general a symmetric distribution is not predicted for all explosion parameters. We note that we have also tested narrower and broader distributions (width $\sim$0.0001 -- 0.1~\mass{}) and find the exact distribution does not have a significant impact on the model observables and does not affect our conclusions. The remaining material in the core is filled in with intermediate mass elements (IMEs).

\begin{figure}
\centering
\includegraphics[width=\columnwidth]{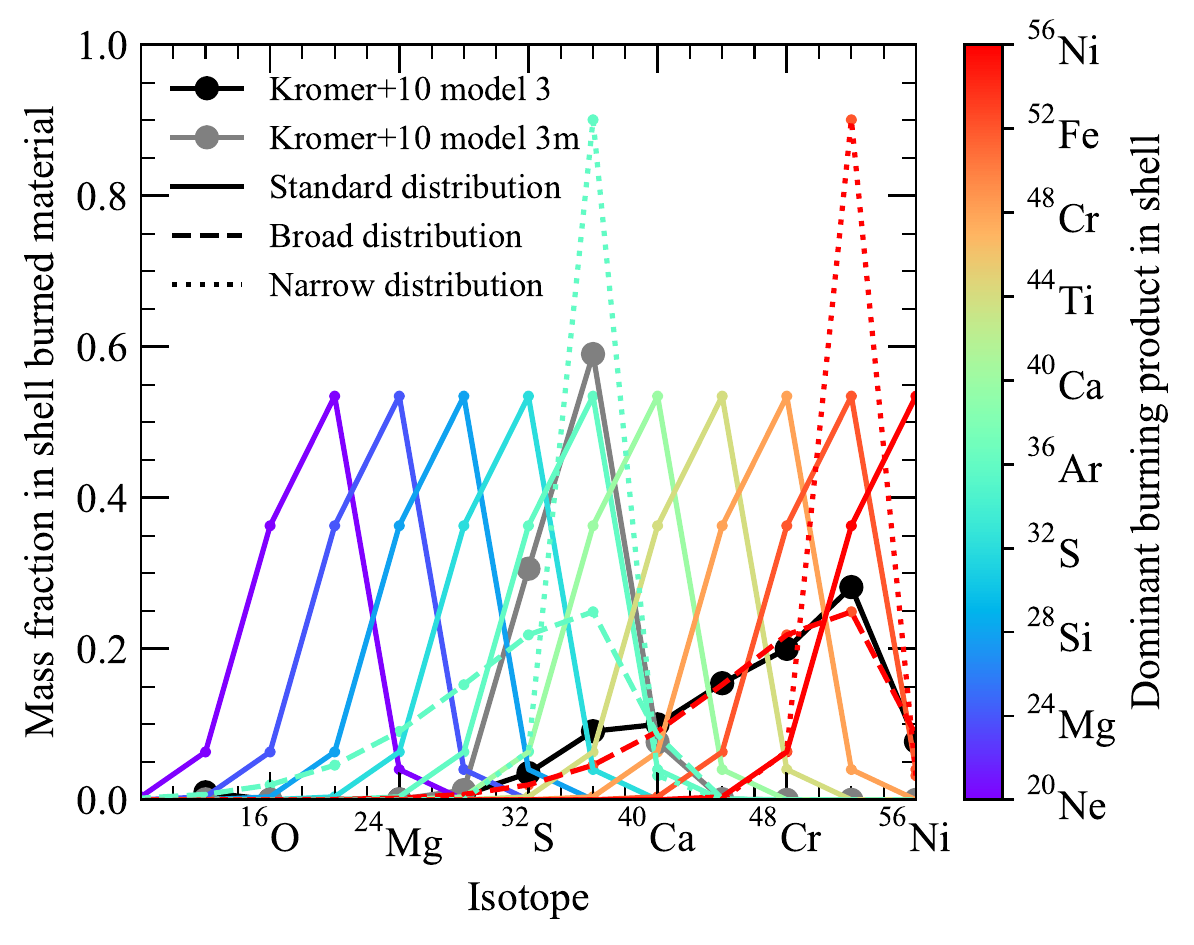}
\caption{Mass fractions of isotopes along the $\alpha$-chain produced in the shell. The relative abundances are shown for the \citet{kromer--10} models assuming a pure helium shell (model 3; black) and a shell that has been polluted with 34\% carbon pre-explosion (model 3m; grey). Coloured lines show the mass fractions of all isotopes, assuming burning progresses to a specific point along the $\alpha$-chain -- given by the colour. Solid lines show our standard isotope distributions, based on model 3m. The dashed line shows a broad distribution designed to mimic that of model 3 from \citet{kromer--10}, while the dotted line shows a narrow distribution in which more mass is burned to the dominant shell product. }
\label{fig:shell_products}
\centering
\end{figure}

\par 
\subsection{Composition of the shell}
\label{sect:construct_shell_comp}
The composition of the helium shell following the explosion remains one of the uncertain properties of double detonation explosions. The goal of this work is to present models covering a large parameter space and systematically investigate differences in observables that result from various assumptions about the helium shell. In Fig.~\ref{fig:ni_masses}(b) we show the fraction of the helium shell that is burned following the explosion (i.e. converted from helium into heavier elements) for a selection of model sets from the literature. It is clear that there can be a large spread in how much of the shell is consumed, depending on different assumptions made within the models (such as when and how ignition is triggered). To investigate the impact of this on the observables, we choose fractions that bracket those predicted by the different explosion models. Specifically, for each total mass we calculate models for which 20\%, 50\%, and 80\% of the helium shell is burned to other elements. These fractions are shown as black dashed lines in Fig.~\ref{fig:ni_masses}(b), while the individual models calculated in this work are shown as black points.

\par

Aside from simply investigating how much of the shell is burned, we also aim to demonstrate the effects of elements that are produced during shell burning. This is strongly dependent on the initial composition of the shell. \citet{kromer--10} present a model in which the helium shell is polluted and contains 34\% \carbon{} (model 3m). This is shown in Fig.~\ref{fig:shell_products}, along with the composition of the unpolluted model (model 3). The choice of 34\% was specifically made to create a helium shell that is mostly burned to \argon{}, which does not produce strong spectroscopic features. As discussed in other studies (e.g. \citealt{shen--09,waldman--11, gronow--20}) the presence of carbon has an important role to play in regulating helium burning and shaping the nucleosynthetic yields of the helium shell. Therefore, the final composition of the helium shell could be tuned by varying the level of pollution before explosion \citep{waldman--11}. \citet{piro--15} has demonstrated that a wide range of carbon pollution fractions could indeed be achieved in the helium shell, depending on specifics of the binary system.

\par

This picture is complicated further however, by the presence of other isotopes besides \carbon{}. In particular, \citet{townsley--19} have shown that $\alpha$-chain burning can stall at much lower pollution fractions ($\sim$11\%) when including \carbon{}, \nitrogen{}, and \oxygen{}. It is clear that the nucleosynthetic yields of the helium shell could show significant variations following explosion. Linking these to specific compositions pre-explosion is a challenging prospect. Therefore, in  this study we explore a wide variety of options and assume that burning in the helium shell could stall at any point along the $\alpha$-chain. We make no claims about specific pre-explosion compositions that could produce such yields. In the following, we refer to the point at which burning stalls as the dominant product in the shell. 

\par

We calculate models for dominant shell products ranging from \sulphur{} to \nickel{}. In our standard model distribution, the relative abundances of other isotopes along the $\alpha$-chain are taken following from the 3m model of \citet{kromer--10}. We chose model 3m for our standard isotope distribution as it represents an intermediate case to the other distributions explored in this work. In addition, \citet{kromer--10} present abundances for each isotope produced in the helium shell. Although \argon{} is the dominant shell product in this model, some amount amount of other isotopes are produced above and below \argon ~in the chain. This is demonstrated in Fig.~\ref{fig:shell_products}, which shows that \argon{} is produced with a mass fraction of $\sim$60\% while the previous $\alpha$-chain isotope (\sulphur{}) has a mass fraction of $\sim$30\% and the next isotope (\calcium{}) has a mass fraction of $\sim$8\%. We use a skew normal distribution that approximates the \citet{kromer--10} 3m distribution in order to determine the relative fraction of all $\alpha$-chain isotopes, for a given dominant shell product. Explosion models and detailed predicted yields covering a range pollution fractions within the helium shell are currently unavailable. Therefore, assuming that some amount of isotopes above and below the dominant shell product are also produced seems reasonable. Taking a functional form similar to an existing explosion model is a pragmatic choice, but we note that the exact quantities are unclear. 

\par

In Fig.~\ref{fig:compositions}(c), we verify that our parameterised approach produces spectra comparable to \citet{kromer--10}. We show a comparison between the maximum light spectrum of model 3 (core mass of 1.025~\mass{}, shell mass of 0.055~\mass{}) and two of our models with similar parameters (core mass of 1.0~\mass{}, shell masses of 0.04 and 0.07~\mass{}). In general, our models show similar results, however the velocities are typically too high. The purpose of Fig.~\ref{fig:compositions}(c) is to demonstrate that our parameterised description of the ejecta is not a limiting factor for the method used here. As discussed in \citet{magee--18}, differences in the radiative transfer code used here (TURTLS) and that of \citet{kromer--10} \cite[ARTIS; ][]{artis} can lead to different observables. This is reflected in the spectra for our models, which are generally bluer than those of \citet{kromer--10}. In addition, the differences in the density profile will have some impact and a combination of these factors appears to result in a shift of features to higher velocities. We again stress that our models are not intended to be reproductions of existing model sets, but are designed to explore a large parameter space. As previously mentioned, by adopting a similar structure to the models of \citet{magee--20} and \citet{magee--20b}, we allow for a direct comparison with models from these works, which were all calculated with the same radiative transfer code.

\par
\subsection{Alternative abundances in the helium shell}

In an effort to quantify the significance of our choice for the relative abundances of isotopes, we calculate two additional sets of models. Firstly, we consider a broad distribution similar to that found for model 3 of \citet{kromer--10}. This corresponds to a higher mass fraction of other isotopes relative to the dominant product in the shell. We also consider a narrow distribution in which the mass fractions of all other isotopes decreases relative to the dominant product. Both cases are shown in Fig.~\ref{fig:shell_products} as a dashed and dotted line for an \iron{} and \argon{} dominated shell, which are the dominant products produced in the standard model 3 and model 3m of \citet{kromer--10}, respectively. 

\par

Together these two sets of models serve to bracket the distributions assumed throughout this work. The effects of these different compositions are discussed further in the appendix in Sect.~\ref{sect:shell_distribution}, but we note that in general the differences are relatively minor.

%

\section{Effects of post-explosion helium shell composition}
\label{sect:shell_composition}

\begin{figure}
\centering
\includegraphics[width=\columnwidth]{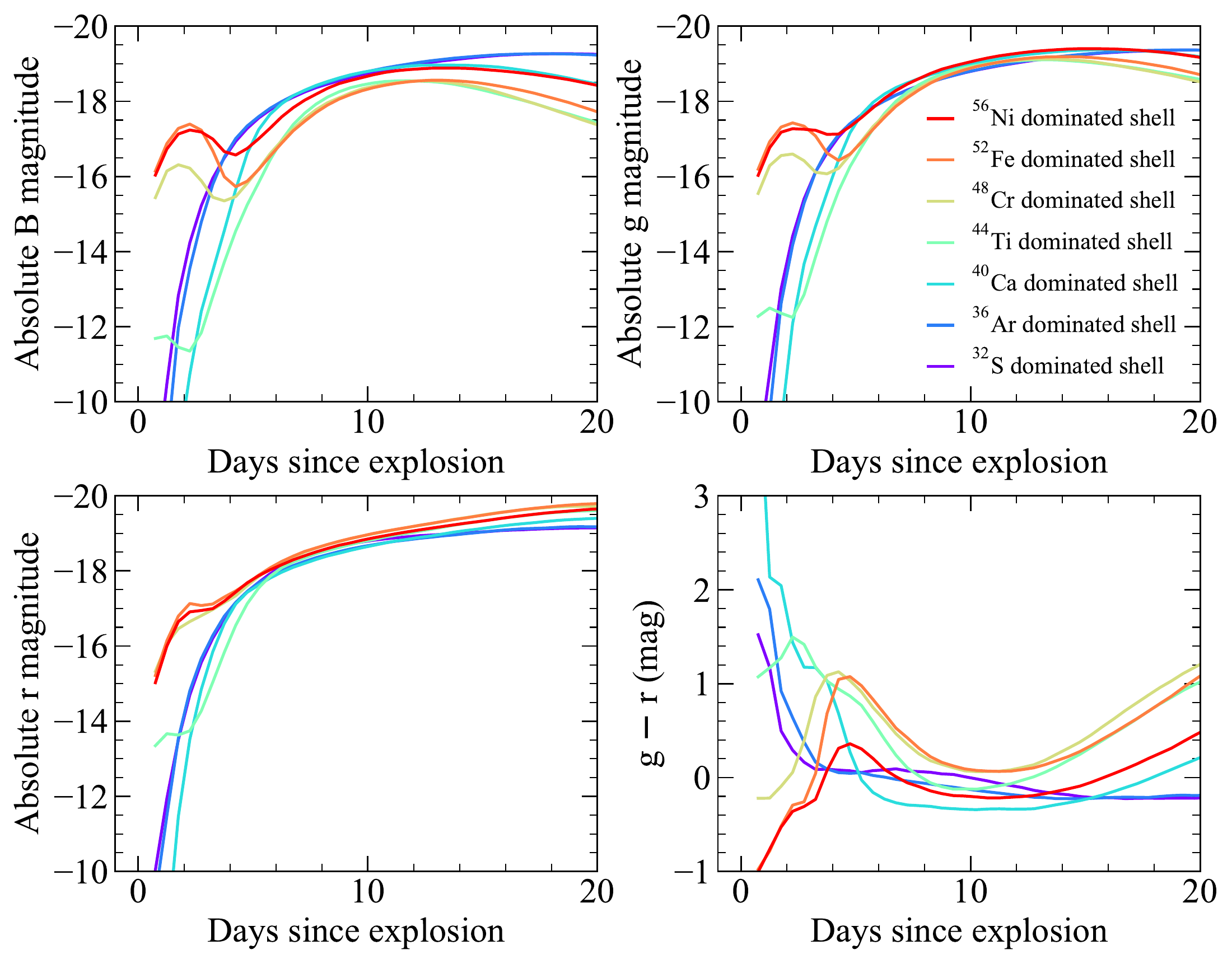}
\caption{Light curves and colours for models with different shell compositions. All models shown have a 1.0~\mass{} core and a 0.07~\mass{} shell, of which 50\% is burned to elements heavier than helium. The dominant $\alpha$-chain product produced in the shell is given by the colours. The relative fractions of all other isotopes in the shell are given following from Fig.~\ref{fig:shell_products}.
}
\label{fig:shell_compositions_lc}
\centering
\end{figure}

In the following section we discuss the results of our radiative transfer modelling. We demonstrate the significant impact of the helium-shell composition on the model light curves and spectra. We compare models with the same core (1.0~\mass{}) and shell (0.07~\mass{}) masses, but different shell compositions for our standard isotope distribution (Sect.~\ref{sect:construct_shell_comp}, Fig.~\ref{fig:shell_products}). For this comparison of the effect of different dominant products in the helium shell, we focus on the models in which 50\% of the helium shell is burned to heavier elements. Other models within our set show similar variations for different shell compositions.

\par

\subsection{Light curves} 
Figure~\ref{fig:shell_compositions_lc} shows the effect of the shell composition on the light curve and colour evolution. Similar to previous studies, we find that those models with $\alpha$-chain burning progressing to IGEs (\titanium{} -- \nickel{}), which therefore have relatively large amounts of short-lived radioactive isotopes in their shells (\nickel{}, \iron{}, and \chromium{}), display prominent bumps in their light curves within the days following explosion. Although these bumps are most pronounced at shorter wavelengths, they are also clearly seen in redder filters (e.g. $r$-band). Aside from the shape of the light curve, models with IGE-dominated shells also show a distinct colour inversion. The colours are initially blue and quickly reach a peak red colour within a few days of explosion. At this point the colour evolution turns over and the models become somewhat bluer again, before again turning over and becoming progressively redder towards maximum light. 

\par

For those models in which the shell is dominated by IMEs (\sulphur{} or \argon{}), no early bump is observed due to the lack of the additional radioactive material. Instead, these models show a smooth rise to maximum light, as well as broader and brighter $B$-band light curves than our IGE-dominated shell models. Our IME-dominated shell models also show a relatively flat colour evolution beginning approximately five days after explosion. We have also calculated models for which the assumed $\alpha$-chain burning stalls earlier than \sulphur{} (\neon{}, \magnesium{}, and \silicon{}), however these models are very similar to each other and the \sulphur{}-dominated model. Therefore, our models show that provided the initial composition of the helium shell is such that burning stops at IMEs, the relative abundances of this burned material are generally unimportant for shaping the evolution of the observables further. 

\par

Interestingly, our \calcium{}-dominated model represents an intermediate case between the IME- and IGE-dominated shells. No early light curve bump is observed and the $B$-band in particular shows a longer dark phase (i.e. the time between explosion and the first light emerging from the supernova) than all other models. At the same time, the colour evolution does not show an inversion similar to the IGE-dominated shells, but is significantly redder at maximum light compared to the IME-dominated shells. Although \calcium{} is the dominant product in the shell ($\sim$55\% of the burned material), a small amount of \titanium{} is also present ($\sim$5\% of the burned material). The additional opacity contribution from \titanium{} will act to more effectively blanket the blue flux than in the other IME-dominated shell models, which do not contain \titanium{}. On the other hand, the \calcium{}-dominated model also lacks a contribution from any radioactive material in the shell, as in the case of the IGE-dominated shell models. Together, both of these properties will cause the lack of additional flux at early times and the redder colours at later times. 

\par
\subsection{Spectra}

\begin{figure}
\centering
\includegraphics[width=\columnwidth]{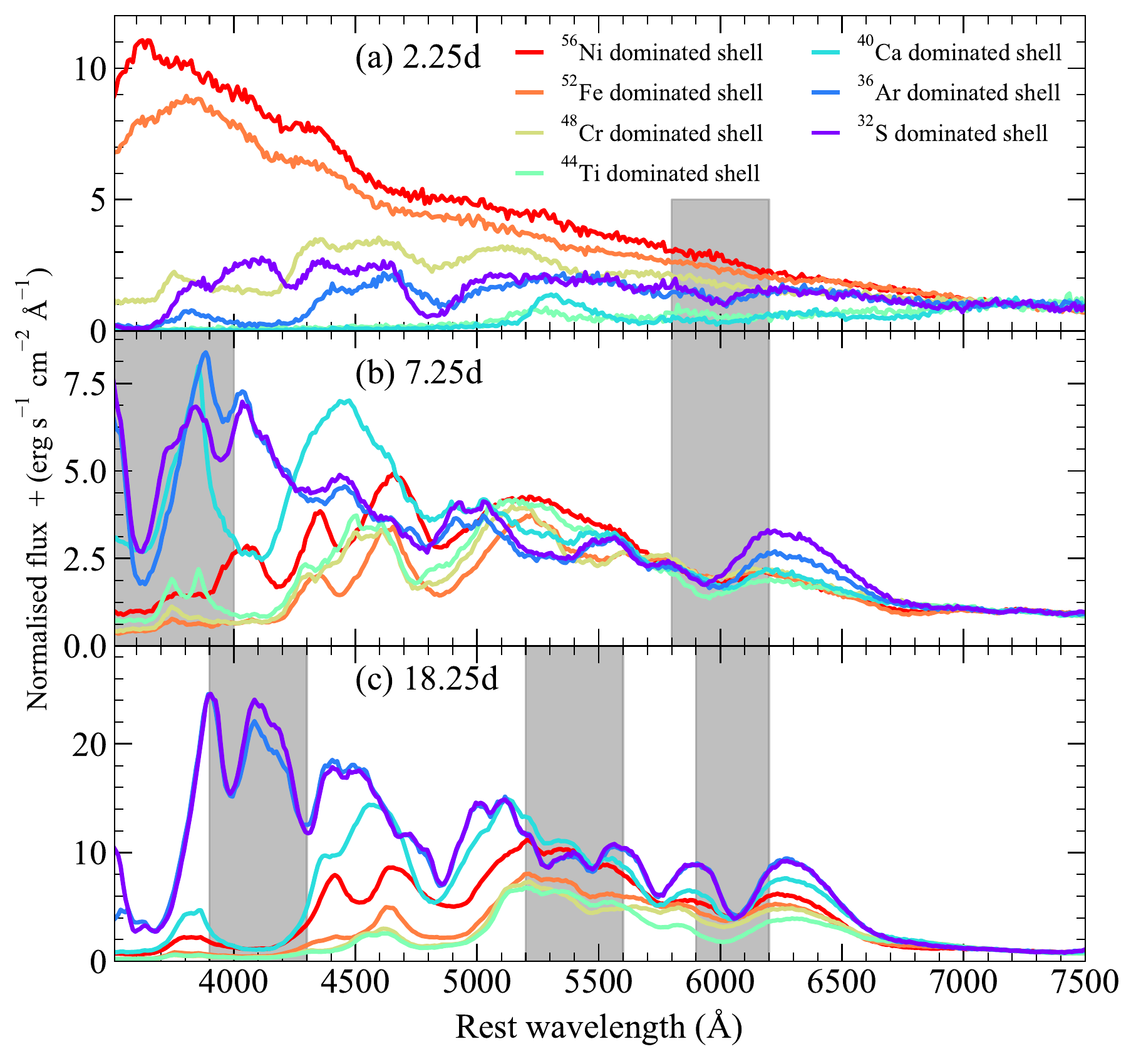}
\caption{Spectra for models with different shell compositions. All models shown have a 1.0~\mass{} core and a 0.07~\mass{} shell, of which 50\% is burned to elements heavier than helium. The dominant $\alpha$-chain product produced in the shell is given by the colours. The relative fractions of all other isotopes in the shell are given following from Fig.~\ref{fig:shell_products}. Spectra are shown at three epochs: 2.25\,d, 7.25\,d, and 18.25\,d after explosion. All spectra are normalised to the flux between 7\,000 -- 7\,500~\AA. Features discussed in the main text are shown as shaded regions. 
}
\label{fig:shell_compositions_spectra}
\centering
\end{figure}

In Fig.~\ref{fig:shell_compositions_spectra}, we show the spectral evolution of our models with different shell compositions. Spectra are shown at 2.25\,d, 7.25\,d, and 18.25\,d after explosion. At 2.25\,d after explosion, our models dominated by \nickel{} and \iron{} are substantially bluer than all other models and show relatively featureless spectra. Despite still containing short-lived isotopes near the surface of the ejecta, Fig.~\ref{fig:shell_compositions_spectra} shows that our \chromium{}-dominated model spectrum is much redder than either the \nickel{}- or \iron{}-dominated model spectra. As shown in Fig.~\ref{fig:noebauer_comp}, \iron{} is the dominant source of luminosity for the early light curve bump -- due to its short half-life. Although some \iron{} is present in the shell of our \chromium{}-dominated model, it has a much lower fraction than in the \nickel{}- or \iron{}-dominated models -- hence there is a lower luminosity and less heating, producing a fainter and redder spectrum during the early bump. Our \chromium{}-dominated spectrum also shows a strong absorption feature due to \ion{S}{ii} at $\sim$4\,800~\AA. At 2.25\,d, our IME-dominated models are still in the dark phase (i.e. very little luminosity has actually escaped). Despite their low luminosity, a weak \ion{Si}{ii}~$\lambda$6\,355 feature is still visible in our \argon{}- and \sulphur{}-dominated models, as well as the \ion{S}{ii} feature around $\sim$4\,800~\AA\, that is also visible in our \chromium{}-dominated model.

\par

One week after explosion, the \nickel{}- and \chromium{}-dominated shell models have become significantly redder. Much of the flux below $\lesssim$4\,000~\AA\, has been blanketed out in all models with IGE-dominated shells. These models also show a broad absorption feature due to \ion{Ti}{ii} around $\sim$4\,200~\AA\, (with the exception of the \nickel{}-dominated model, which does not contain Ti in the shell). Our \chromium{} model shows remarkably little spectral evolution between the two epochs presented here relative to other models within our set. In the IGE-dominated models, a few additional features are produced at longer wavelengths (most notably the \ion{Si}{ii}~$\lambda$6\,355 feature), but in general they are weaker and broader than in models that do not contain IGE in the shell. For our \argon{}- and \sulphur{}-dominated models, the spectra are now considerably bluer than the IGE-dominated models. Absorption profiles due to IME, such as \ion{Si}{ii}~$\lambda$6\,355 and the \ion{S}{ii}-W feature can be observed. Both models also show strong \ion{Ca}{ii} absorption at $\sim$3\,600~\AA.
\par

Moving to maximum light, more of the blue flux in our IGE-dominated models has been blanketed out. At this epoch, the spectra show very little flux below $\sim$4\,300~\AA\, and again show a broad, flat \ion{Ti}{ii} profile between $\sim$3\,900 -- 4\,300~\AA. Around $\sim$4\,700 -- 4\,900~\AA, the IGE-dominated models show a similarly broad and flat feature due to \ion{Cr}{ii} and \ion{Ti}{ii}. For these models, features due to IME (\ion{Si}{ii}~$\lambda$6\,355, \ion{Si}{ii}~$\lambda$5\,972, and \ion{S}{ii}-W) are again broader and weaker than in the IME-dominated shell models. 

\par

\begin{figure}
\centering
\includegraphics[width=\columnwidth]{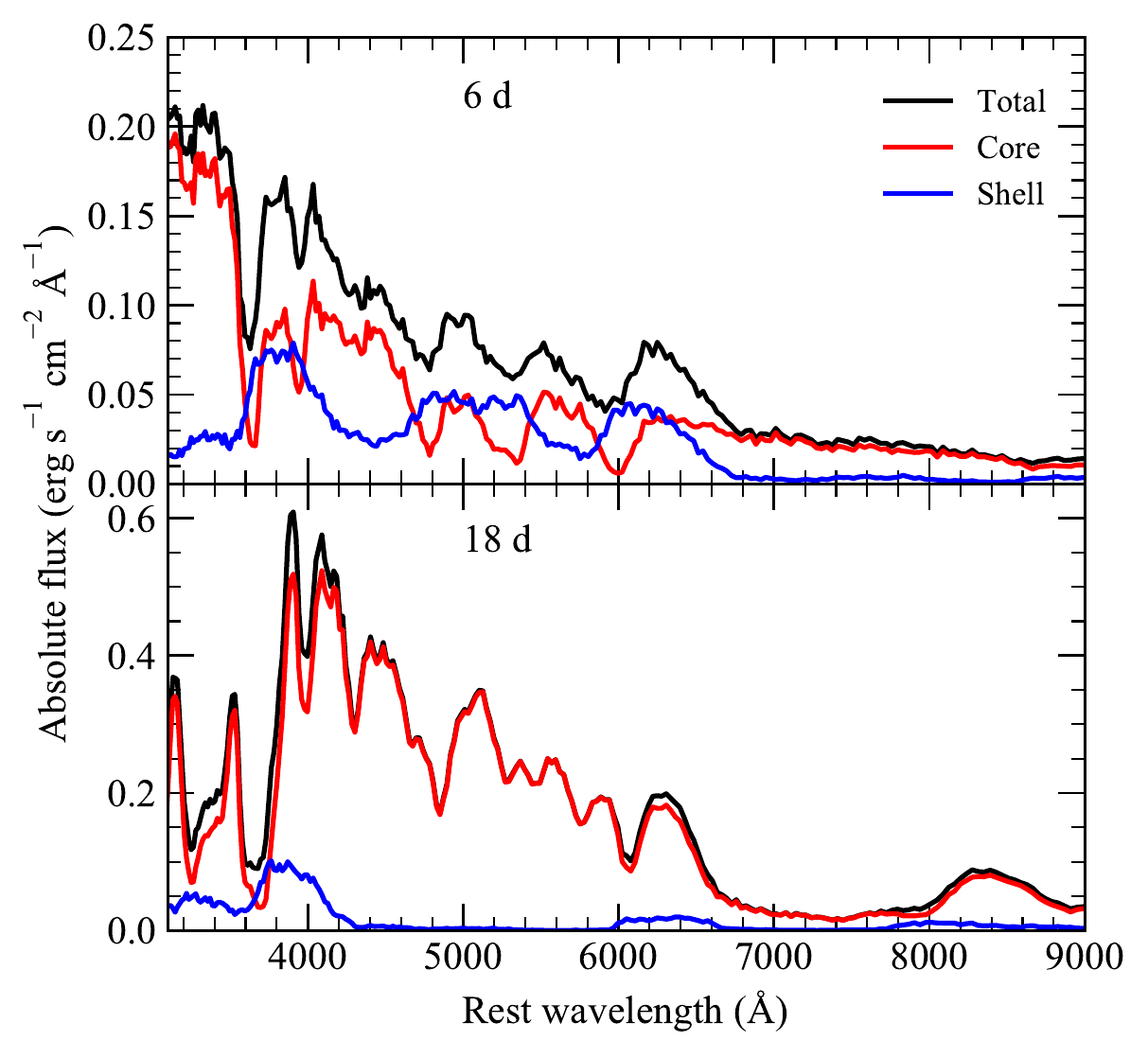}
\caption{Contribution of material in the helium shell and core to the observed spectra at different phases. Spectra are calculated by binning packets separately, depending on the location of their last interaction.
}
\label{fig:hvfs}
\centering
\end{figure}

High-velocity features have been reported in a number of SNe~Ia at early times (e.g. \citealt{childress--14, maguire--14, zhao--15}). The origin of these features is unclear, but one proposed scenario is from an abundance or density enhancement that may be due to circumstellar interaction or intrinsic to SN itself \citep{mazzali--05, tanaka--08}. Double detonation models producing IMEs in the helium shell would be a natural method producing such an abundance enhancement. To investigate whether our IME-dominated shells produce similar features and if these can be attributed to high velocity material in the shell, we calculate the contribution of the shell material to the synthetic spectra. During the simulation, we track the location at which a Monte Carlo packet experiences its last interaction. In Fig.~\ref{fig:hvfs}, we show separate spectra produced by binning Monte Carlo packets that last interacted with material in either the shell or the core. We note that we are only able to track the location of real packets (rather than virtual packets; see \citealt{magee--20b}), therefore the signal-to-noise ratio of these spectra is lower than others presented throughout this work. Nevertheless, Fig.~\ref{fig:hvfs} shows that within the first approximately one week since explosion, the shell material does contribute to the production of high velocity features. Specifically, the \ion{Si}{ii}~$\lambda$6\,355 feature is shifted to higher velocities and broadened when including contributions from the shell. Around maximum light however, there is a negligible impact from the shell material. While our models indicate that helium shell ash could provide one explanation for high velocity features seen in some SNe~Ia at early times, further modelling work is required to fully constrain the abundance profiles required.

\par
\subsection{Summary 
}

Our models clearly show the impact of the post-explosion helium shell composition on the observables. Those models in which the shell is burned mainly to IGE show an early bump in the light curve, a colour inversion, and significantly reddened spectra from approximately one week after explosion. Conversely, our models in which the shell mostly contains IMEs do not show an early bump and instead have a relatively flat colour evolution. In addition, we find that as long as burning within the shell does not progress to IGEs, the model observables show much smaller variations than those for which the shell is dominated by IGEs. This would indicate that meticulous fine-tuning is not necessary to avoid the impact of the helium shell ash on the observables -- provided burning ceases at a certain point, the exact composition of the shell is mostly irrelevant.

%

%

\section{Effects of the burned mass}
\label{sect:burned_mass}

In this section, we demonstrate how the amount of burned material above the core affects the model observables. To focus our discussion, we limit our comparisons to models with a core mass of 1.0~\mass{}. Our models are controlled by both the mass of helium shell and the amount of the shell that is assumed to be burned during the explosion. As the mass of the helium shell also determines the amount of \nickel{} produced during the explosion, which will have a significant impact on the model observables, it is not possible to explore solely the effect of the total amount of material burned. Therefore, in Sect.~\ref{sect:shell_mass} we discuss the effects of the helium shell mass and in  Sect.~\ref{sect:burned_fraction} we discuss the role of the burned fraction.

\subsection{Impact of helium shell mass on light curves and spectra}
\label{sect:shell_mass}

\begin{figure}
\centering
\includegraphics[width=\columnwidth]{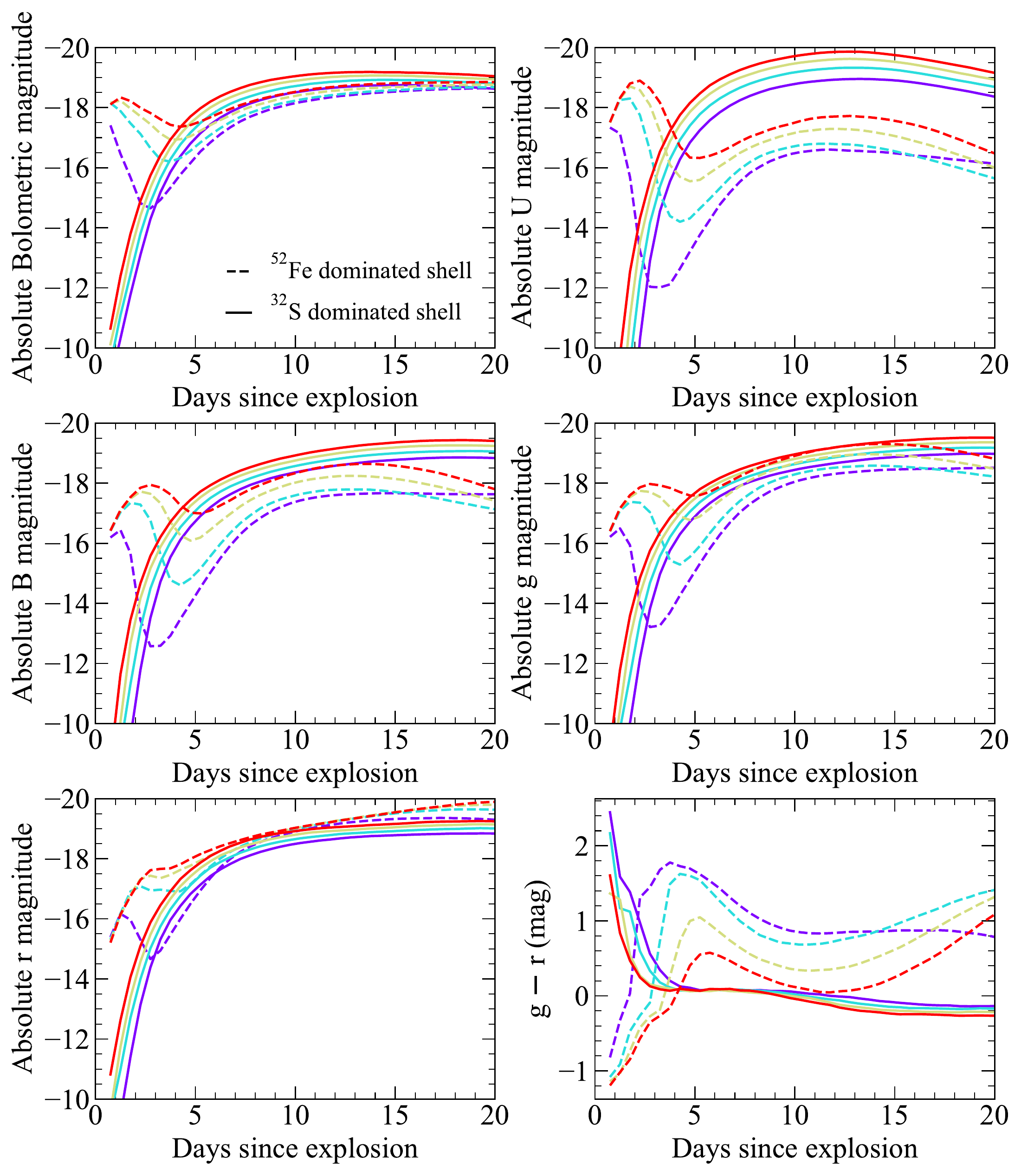}
\caption{Light curves and colours for models with different shell masses. All models shown have a 1.0~\mass{} core and
we assume 80\% of the shell is burned to elements heavier than helium. We show two representative cases in which the composition of the shell is dominated by either \sulphur{} or \iron{}.
}
\label{fig:shell_mass_lc}
\centering
\end{figure}

\par

\begin{figure*}
\centering
\includegraphics[width=\textwidth]{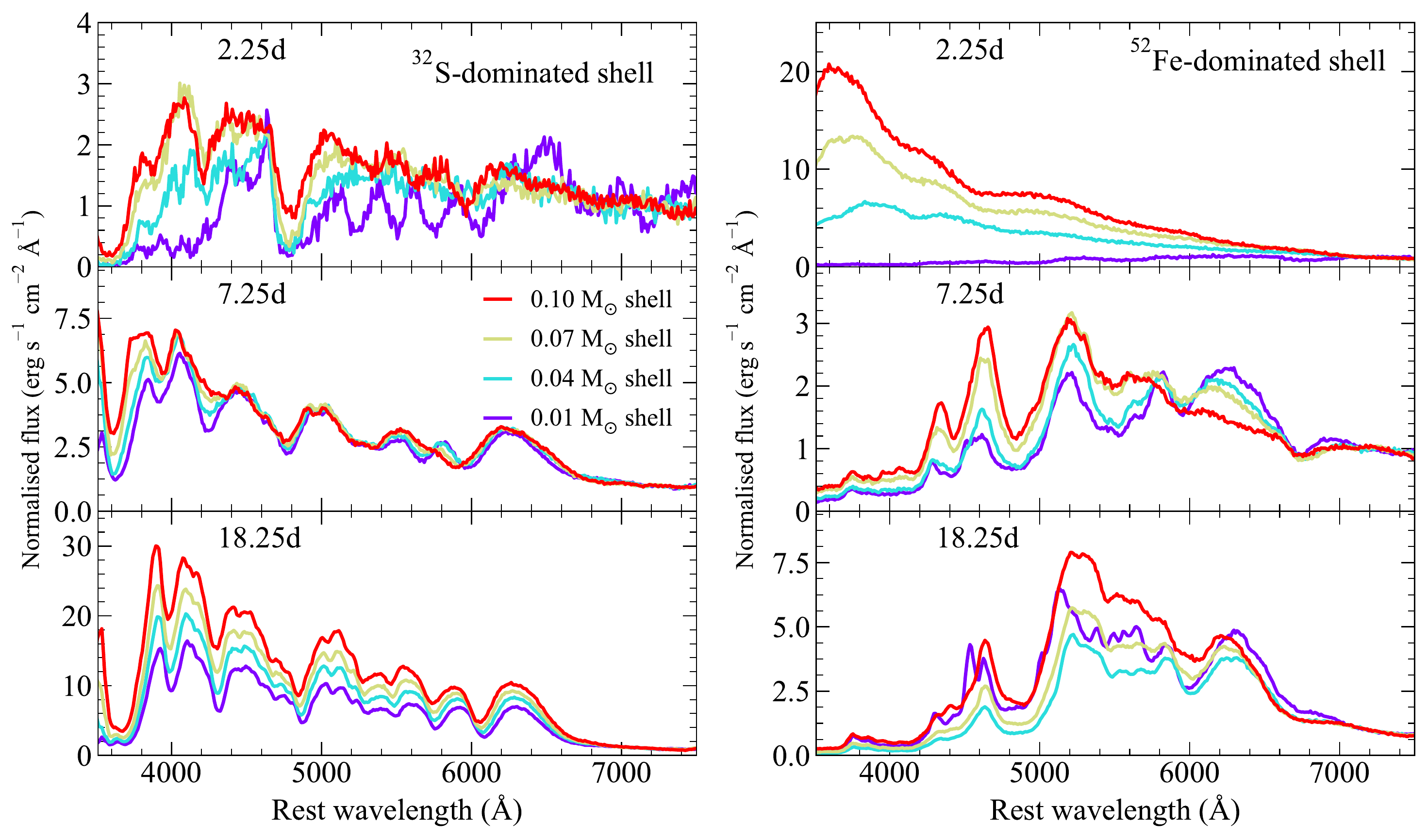}
\caption{Spectra for models with different shell masses. All models shown have a 1.0~\mass{} core and
we assume 80\% of the shell is burned to elements heavier than helium. We show two representative cases in which the composition of the shell is dominated by either \sulphur{} or \iron{}. Spectra are shown at three epochs: 2.25\,d, 7.25\,d, 18.25\,d after explosion. All spectra are normalised to the flux between 7\,000 -- 7\,500~\AA. 
}
\label{fig:shell_mass_spectra}
\centering
\end{figure*}

In Fig.~\ref{fig:shell_mass_lc} we show the light curve and colour evolution for models with varying shell masses, while the spectral evolution is shown in Fig.~\ref{fig:shell_mass_spectra}. We limit our comparison to the \iron{}- and \sulphur{}-dominated shells, which are representative of trends observed for IGE- and IME-dominated shells, respectively (see Sect.~\ref{sect:shell_composition}).

\par

As discussed in Sect.~\ref{sect:shell_composition}, our \sulphur{}-dominated shell models do not produce an early bump in the light curve, but there is still some variation among the different shell masses. This is not primarily driven by material in the shell, but rather the different \nickel{} masses in the core of the white dwarf (Sect.~\ref{sect:construct_core_comp}, Fig.~\ref{fig:ni_masses}). Therefore, models with more massive shells are brighter and somewhat bluer simply due to the increased ejecta mass and hence \nickel{} mass.  Figure~\ref{fig:shell_mass_spectra} shows that these models also produce similar spectra, with the primary differences being the luminosity and colour. Differences between the spectra of models with different shell masses are most pronounced at early times, where lower mass shells show stronger  \ion{Si}{ii} and \ion{S}{ii} features, likely due to their lower temperatures. For our IME-dominated models we also note there is also a degeneracy between the core and shell masses. For the models presented here, provided the total ejecta mass is the same, the distinction between the core and shell is unimportant. For example our \sulphur{}-dominated model with a 1.0~\mass{} core and 0.1~\mass{} shell and model with a 1.1~\mass{} core and 0.01~\mass{} shell produce similar light curves and spectra.

\par

For our \iron{}-dominated shells, Fig.~\ref{fig:shell_mass_lc} shows that all models produce a light curve bump. The timescale of the bump varies significantly ($\sim$1 -- 5 days) for the broad-band light curves, with lower mass shells producing shorter-lived and more rapidly evolving bumps. As demonstrated by Fig.~\ref{fig:shell_mass_lc}, this is primarily due to temperature evolution for the different models, as there is significantly smaller spread in the bump timescales in bolometric light. Unlike the \sulphur{}-dominated shell models, the shell mass has a considerable impact on the colour evolution for the \iron{}-dominated models. Smaller shell masses produce a more rapid and extreme change in colour within the first five days after explosion. In addition, beginning approximately 10 days after explosion, the lowest mass shell model (0.01~\mass{}) shows a relatively flat colour evolution towards maximum light. In contrast, models with more massive shells become significantly redder over this same period. The 0.10~\mass{} shell model remains redder than both the 0.07 and 0.04~\mass{} shell models for all times presented here, however the overall difference between their respective colours decreases with time. This likely points to two competing effects -- the influence of line blanketing from the shell and the different \nickel{} masses causing different temperatures. More massive shells will produce more line blanketing and hence one may expect redder colours, but this is not observed for the models presented here. In this case, as the core mass is the same, the increase in the shell mass results in an increased \nickel{} mass that keeps the ejecta hotter and bluer. Figure~\ref{fig:shell_mass_spectra}(d) shows that, at 2.25\,d after explosion, the temperature is the primary difference between the models and few features are present. At later epochs, our models show that larger shell masses produce broader \ion{Si}{ii}~$\lambda$6\,355 features and weaker IME features overall.

\par

\subsection{Impact of the burned fraction percentage on light curves and spectra}
\label{sect:burned_fraction}

The amount of material in the shell converted from helium to heavier elements is also a free parameter of our models. We have investigated burned fractions of 20\%, 50\%, and 80\%, which approximately span the range predicted by different explosion models (see Fig.~\ref{fig:ni_masses}(b)). The differences between these models are fairly straightforward and follow the trends one may expect. For IME-dominated shell models, the burned fraction has no effect on the resultant observables. For IGE-dominated shell models, a higher burned fraction will result in a brighter bump at early times and redder colours at later times.

%

\section{Model rise times and bump timescales}
\label{sect:rises}

\begin{figure*}
\centering
\includegraphics[width=\textwidth]{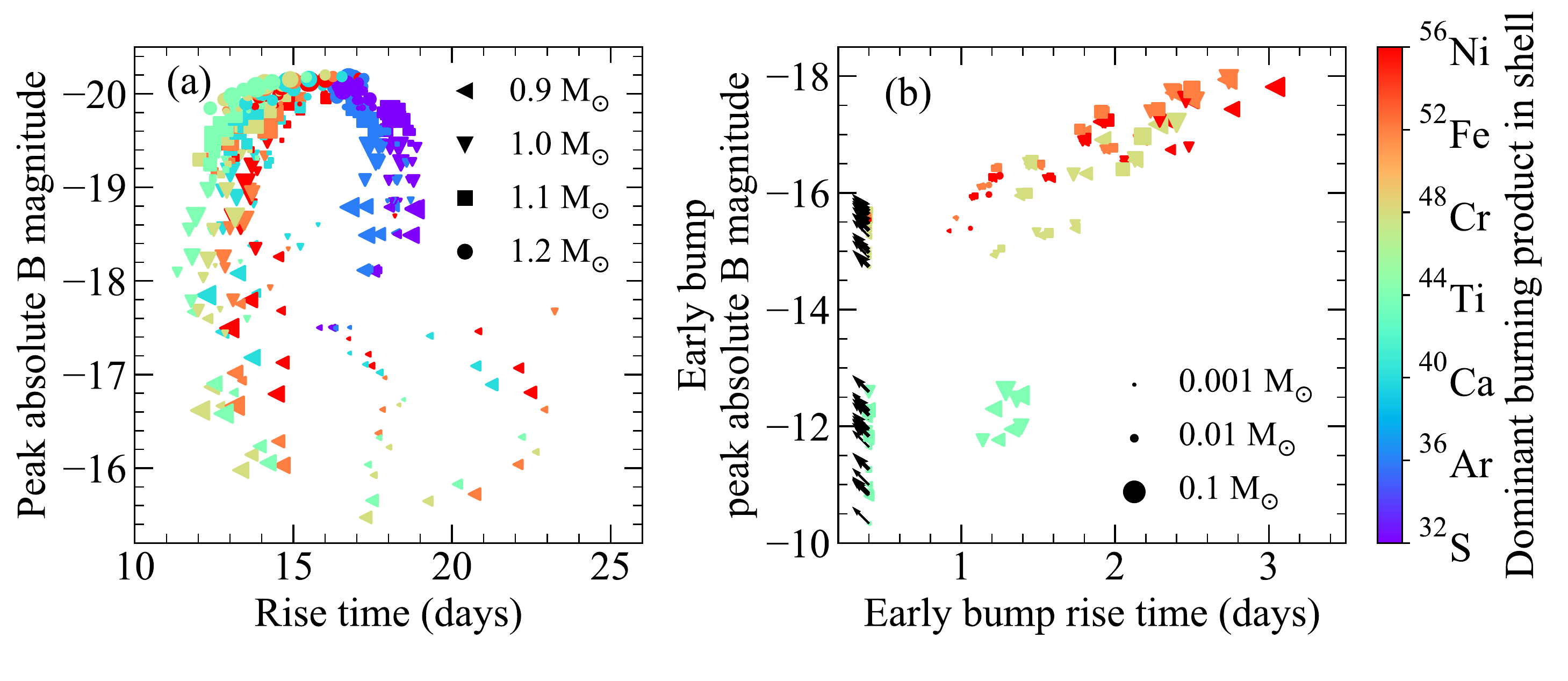}
\caption{ {\it Panel a:} Peak absolute $B$-band magnitudes against rise time to peak $B$-band magnitude. {\it Panel b:} Peak absolute $B$-band magnitudes of the early light curve bump against time to reach the peak of the bump. We note that all models with IME dominated shells and some models with high core masses do not show early bumps and therefore are neglected. For models in which the light curve is already declining at the start of our simulation (0.5\,d), we consider these as upper limits and show them as black arrows. In both panels, each model is coloured based on the dominant element produced in the shell. The size of each point is proportional to the burned mass of the helium shell (i.e. the product of the shell mass and burned fraction), while the shape of each point denotes the mass of the core. 
}
\label{fig:rises}
\centering
\end{figure*}

Here we discuss the rise times and peak magnitudes of the models presented in this work and demonstrate the range of magnitudes and timescales for early light curve bumps. In Fig.~\ref{fig:rises}(a) we show the $B$-band rise times and peak absolute $B$-band magnitudes for our models with the standard isotope distribution. The difference between our IGE- and IME-dominated shell models is readily apparent. We find that, in general, those models with IGE-dominated shells show shorter rises, with a median rise time of 13.8$\pm$2.3\,d, compared to those with IME-dominated shells, which have a median rise time of 17.6$\pm$0.7\,d. The longer rise time of the IME-dominated models is more typical of normal SNe~Ia (e.g. \citealt{ganeshalingam--2011, firth--sneia--rise, miller--20}). Although in general we find that models with IGE-dominated shells show shorter rise times, there are some notable exceptions. For a 0.9~\mass{} core, some models with low-mass shells (0.01 and 0.04~\mass{}) can show longer rise times than similar models with higher mass cores. In these cases, the longer rise times result from a combination of the compact \nickel{} distribution and less extreme line blanketing of the low-mass shell. For our IME-dominated models, the scatter in the peak absolute $B$-band magnitude is driven simply by differences in \nickel{} mass due to the various core and shell masses explored here. Models with IGE-dominated shells however, show a significantly larger scatter ($\gtrsim$4~mag compared to $\sim$2~mag for IME-dominated shells) due to line blanketing from the material in the shell. Indeed at longer wavelengths that are less sensitive to line blanketing (e.g. $r$-band), both the IGE- and IME-dominated shell models show a similar scatter in peak magnitudes (again, due to the differences in the \nickel{} mass), although the IGE-dominated models are systematically brighter as much of the blue flux has been reprocessed to longer wavelengths by the shell. 

\par

Figure~\ref{fig:rises}(b) shows the properties of the early light curve bumps. We calculate the time since explosion to reach the peak of the bump and the magnitude at this point. For some models, the light curve is already declining at the beginning of our simulations (0.5\,d after explosion). We therefore consider these points as limits and show them as black arrows in Fig.~\ref{fig:rises}(b). We do not include models with IME-dominated shells as they do not show a bump at early times. In addition some models, such as those with large total masses ($\gtrsim$1.2~\mass{}), do not show pronounced bumps in their light curves due to their high \nickel{} masses and extended distributions. In other words, there is no clear decline in the light curve within the first few days of explosion. These models are also not included in Fig.~\ref{fig:rises}(b). Among the models shown in Fig.~\ref{fig:rises}(b), there is a general trend that brighter bumps are also typically longer lasting. Models with \titanium{}-dominated shells however, deviate from this trend. Following from Fig.~\ref{fig:shell_products}, in our \titanium{} dominated model only a small amount of \chromium{} is contained within the shell. Therefore this set of models contain a significantly smaller mass of radioactive isotopes in the shell compared to our other IGE dominated models. We also note that the models shown as limits in Fig.~\ref{fig:rises}(b) hint at the possibility of bright and very short lived bumps -- less than 0.5\,d. It is highly likely that such bumps could be missed in most current surveys. 

%

 %
\section{Comparison with a Chandrasekhar-mass model containing a \nickel{} excess}
\label{sect:comparisons_ni}

\citet{magee--20b} present light curves and spectra of Chandrasekhar mass models that contain an excess of \nickel{} (a \nickel{} shell) in the outer ejecta. Qualitatively, these models show similar behaviour (light curve bumps at early times and line blanketing closer to maximum light) to double detonations in which a significant fraction of IGEs is produced in the shell. Here we perform a comparison between these two cases and investigate ways in which they may be distinguished from each other, based on the early light curve bump and spectra at maximum light. 

\par

\begin{figure}
\centering
\includegraphics[width=\columnwidth]{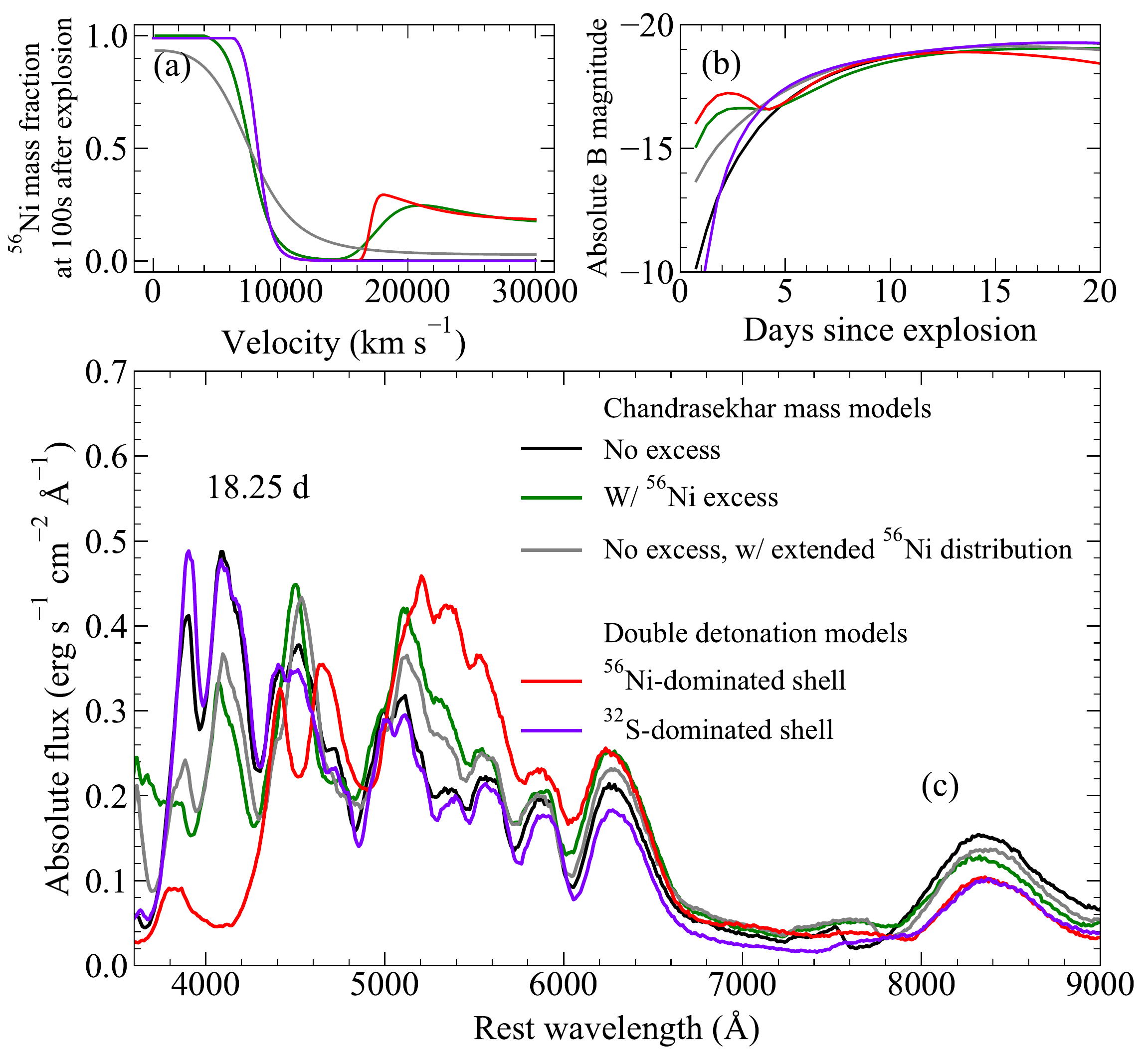}
\caption{{\it Panel a:} Comparison between the \nickel{} distributions for models presented here. We show a Chandrasekhar mass model from \citet{magee--20b} that contains a \nickel{} excess in the outer ejecta and the corresponding model without an excess, in addition to a model with an extended \nickel{} distribution. Double detonation models with a \nickel{}- and \sulphur{}-dominated shell (red and purple, respectively) are also shown. {\it Panel b:} $B$-band light curve for models presented here. {\it Panel c:} Comparison of spectra for our models at 18.25\,d after explosion. 
}
\label{fig:ni_dists_comp}
\centering
\end{figure}

In Fig.~\ref{fig:ni_dists_comp}(a), we show the \nickel{} distributions of Chandrasekhar-mass model with and without a \nickel{} excess compared to the sub-Chandrasekhar double detonation models explored in this work. We show one of the Chandrasekhar mass models from \citet{magee--20b} that does not contain a \nickel{} excess (black in Fig.~\ref{fig:ni_dists_comp}; described as the fiducial SN~2018oh model in \citealt{magee--20b}), along with a model in which a 0.03~\mass{} \nickel{} shell has been added to the outer ejecta (Fig.~\ref{fig:ni_dists_comp}, green). We present an additional Chandrasekhar mass model without a \nickel{} excess, but in which the \nickel{} distribution has been extended, such that \nickel{} is present throughout the ejecta with a mass fraction that decreases monotonically towards the outer ejecta (Fig.~\ref{fig:ni_dists_comp}, grey). For our sub-Chandrasekhar mass double detonation model with a 1.0~\mass{} core and a 0.07~\mass{} shell dominated by \nickel{} (WD1.00\_He0.07\_BF0.50\_DP56Ni; Fig.~\ref{fig:ni_dists_comp}, red), the total \nickel{} mass and \nickel{} distribution in the outer ejecta is similar to the \nickel{} excess model. Finally, we also show the same model with a \sulphur{}-dominated shell (WD1.00\_He0.07\_BF0.50\_DP32S; Fig.~\ref{fig:ni_dists_comp}, blue) as representative of a sub-Chandrasekhar mass model without an excess of \nickel{} in the outer ejecta. We note that the density profile and ejecta mass differs slightly between the double detonation and Chandrasekhar mass models shown here.

Figure~\ref{fig:ni_dists_comp}(b) demonstrates that the sub-Chandrasekhar mass double detonation \nickel{}-dominated shell model shows a more pronounced bump that rises and declines within a few days compared to the more plateau-like shape of the Chandrasekhar mass \nickel{} excess model. Even for double detonation models with a lower \nickel{} mass fraction in the outer ejecta (i.e. different burned fractions), the \nickel{}-dominated shells do not reproduce the shape of the \nickel{} excess models. Such a difference in light curve shape, despite similar \nickel{} distributions, serves to further highlight the importance of the additional radioactive isotopes produced in the double detonation models. For the Chandrasekhar-mass \nickel{} excess model, \nickel{} is the only radioactive isotope considered in the ejecta, while the double detonation model also contains \iron{} and a small amount of \chromium{}. Hence, the bump produced in the light curve of the \nickel{}-dominated shell model is more pronounced due to the presence of \iron{}, $^{52}$Mn, and \chromium{}, which have considerably shorter half-lives compared to \nickel{}. At later epochs, the double detonation model with a \nickel{}-dominated shell becomes significantly redder and fainter than the \nickel{} excess model.  Again this points to important differences in the ejecta composition -- the presence of additional IGEs in the double detonation model will more effectively blanket blue flux than just the \nickel{} decay chain as in the \nickel{} excess model. For our \sulphur{}-dominated shell model, the light curve shows a sharper rise and slightly longer dark phase than the model without a clump. In this case, the \nickel{} distribution is somewhat less extended and shows a more rapid change from \nickel{}-rich to -poor ejecta than the Chandrasekhar mass model without a \nickel{} excess. 

\par

In Fig.~\ref{fig:ni_dists_comp}(c) we show the spectra of all models at 18.25\,d after explosion. At this epoch, our Chandrasekhar mass model without a \nickel{} excess and \sulphur{}-dominated shell model show extremely similar spectra (black and purple lines), with the most noticeable difference being that the double detonation model is marginally bluer. Conversely, the \nickel{}-dominated shell model (red line) is significantly different from all other models, including the Chandrasekhar mass \nickel{} excess model (green line). The flux below $\sim$4\,000~\AA\, is essentially completely removed from the spectrum and redistributed to wavelengths $\gtrsim$5\,000~\AA, which show a significantly higher continuum flux. The \ion{Cr}{ii} and \ion{Ti}{ii} features present in the double detonation model at $\sim$4\,000 -- 5\,000~\AA\, easily distinguish it from the \nickel{} excess model.

\par

Comparing Chandrasekhar mass models with a \nickel{} excess in the outer ejecta and sub-Chandrasekhar mass double detonation models in which a \nickel{}-dominated shell is produced as a result of helium shell burning, we find that the two are easily distinguished despite qualitatively similar behaviour. Although thought to be created via a different mechanism, models with \nickel{} excess can also produce an early light curve bump. The shape of the bump however, more closely resembles a plateau compared to the clear peak in the double detonation models. The significant amount of IGEs produced during helium shell burning leads to extremely red colours -- even more so than the \nickel{} excess models, which also show red colours at maximum. Finally, our IGE-dominated shell models also show shorter rise times than the \nickel{} excess models of \citet{magee--20b}.

%

 %
\section{Comparisons with normal SNe Ia}
\label{sect:comparisons_normal}

\begin{figure*}
    \centering
    \begin{subfigure}[b]{0.48\textwidth}\includegraphics[width=\columnwidth]{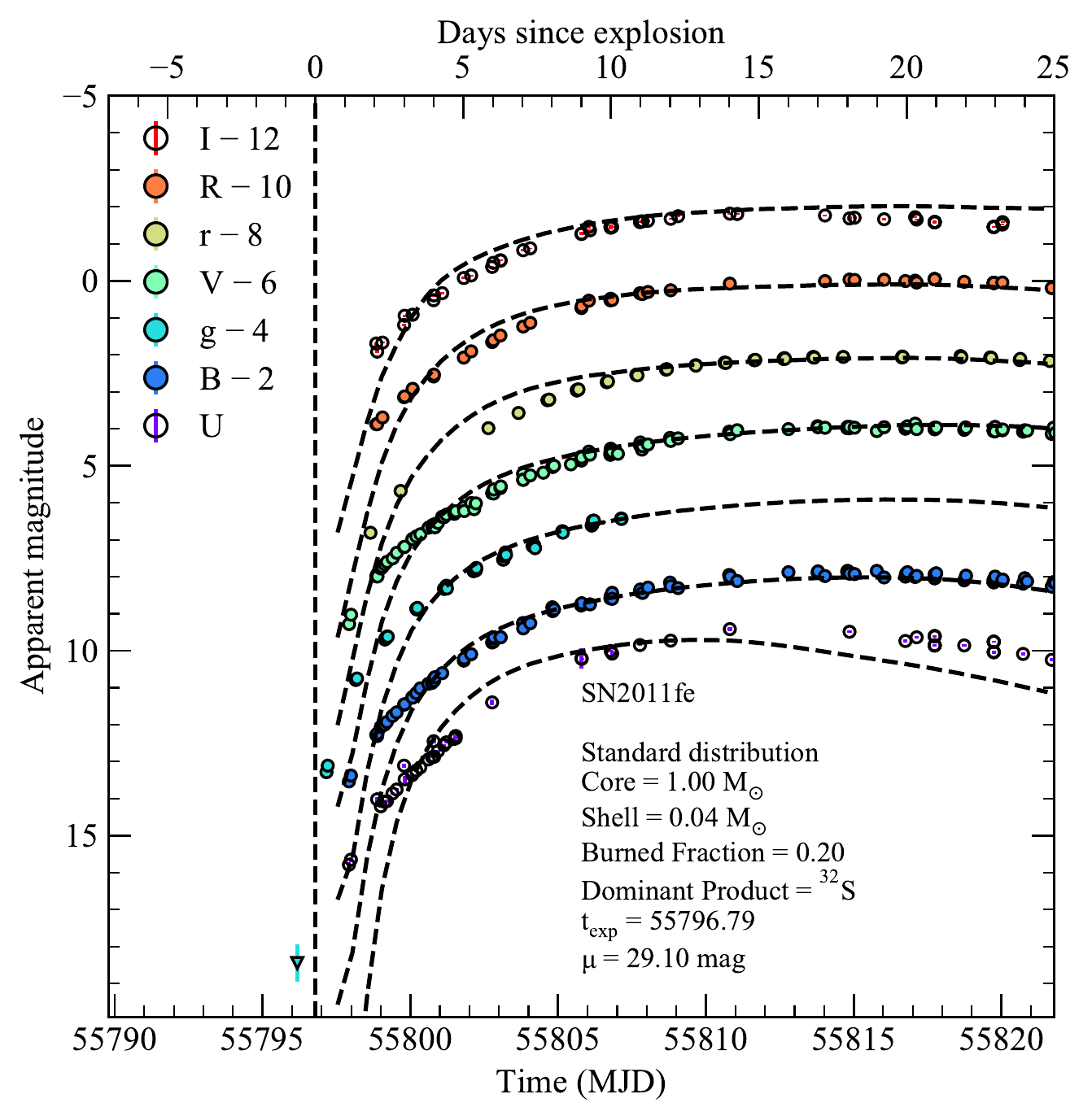}
    \end{subfigure}
    \begin{subfigure}[b]{0.48\textwidth}\includegraphics[width=\columnwidth]{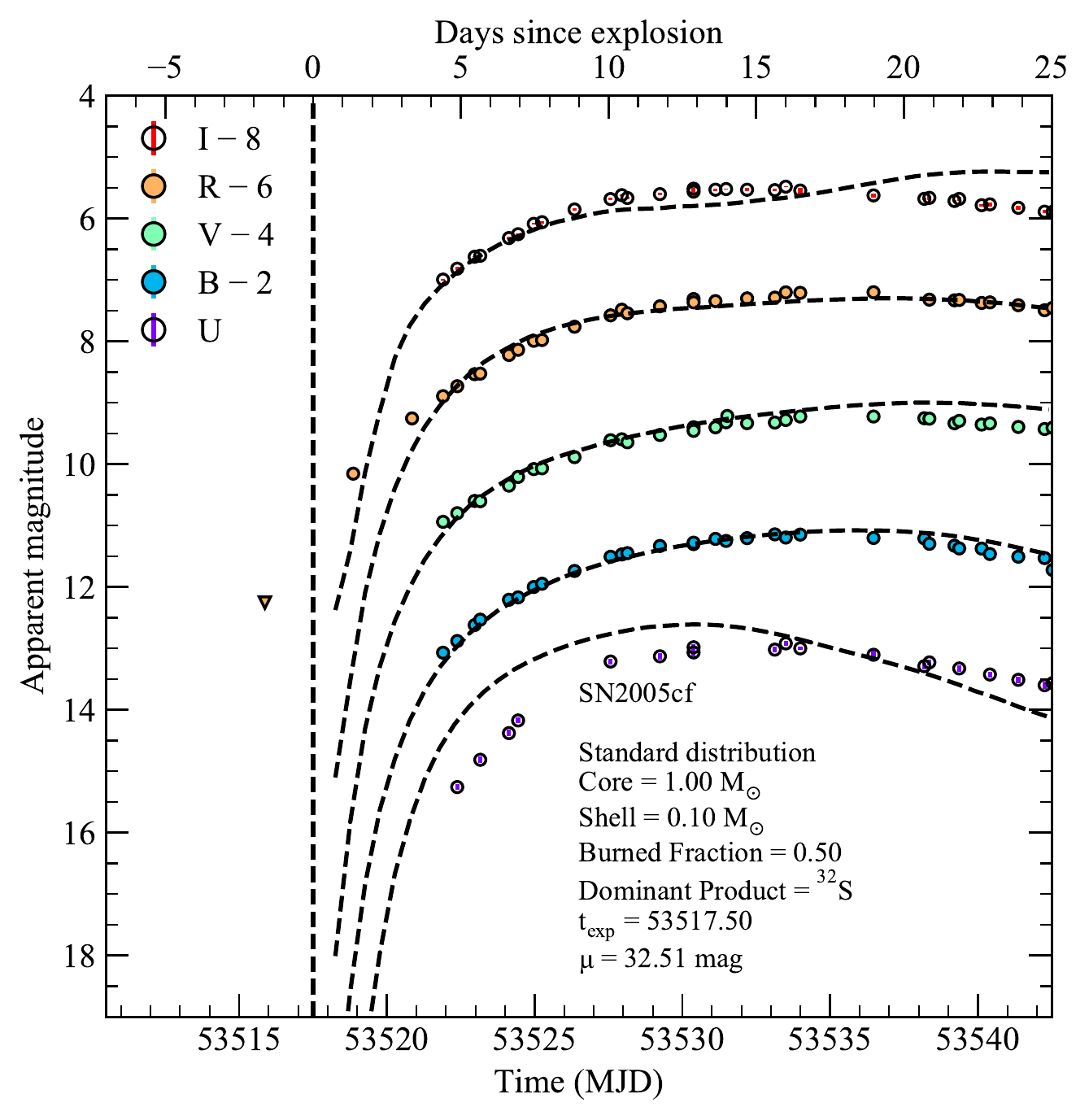}
    \end{subfigure} 
    \caption{Comparisons between the optical light curves of SNe~2011fe and 2005cf (coloured circles) and our sub-Chandrasekhar double detonation model light curves (dashed black lines). The model parameters and assumed distance modulus are given for each object. The estimated time of explosion (based on the comparison with the model light curve) is shown as a vertical dashed line for each SN. 
    }
    \label{fig:normal_lc}
\end{figure*}

\begin{figure*}
    \centering
    \begin{subfigure}[b]{0.48\textwidth}\includegraphics[width=\columnwidth]{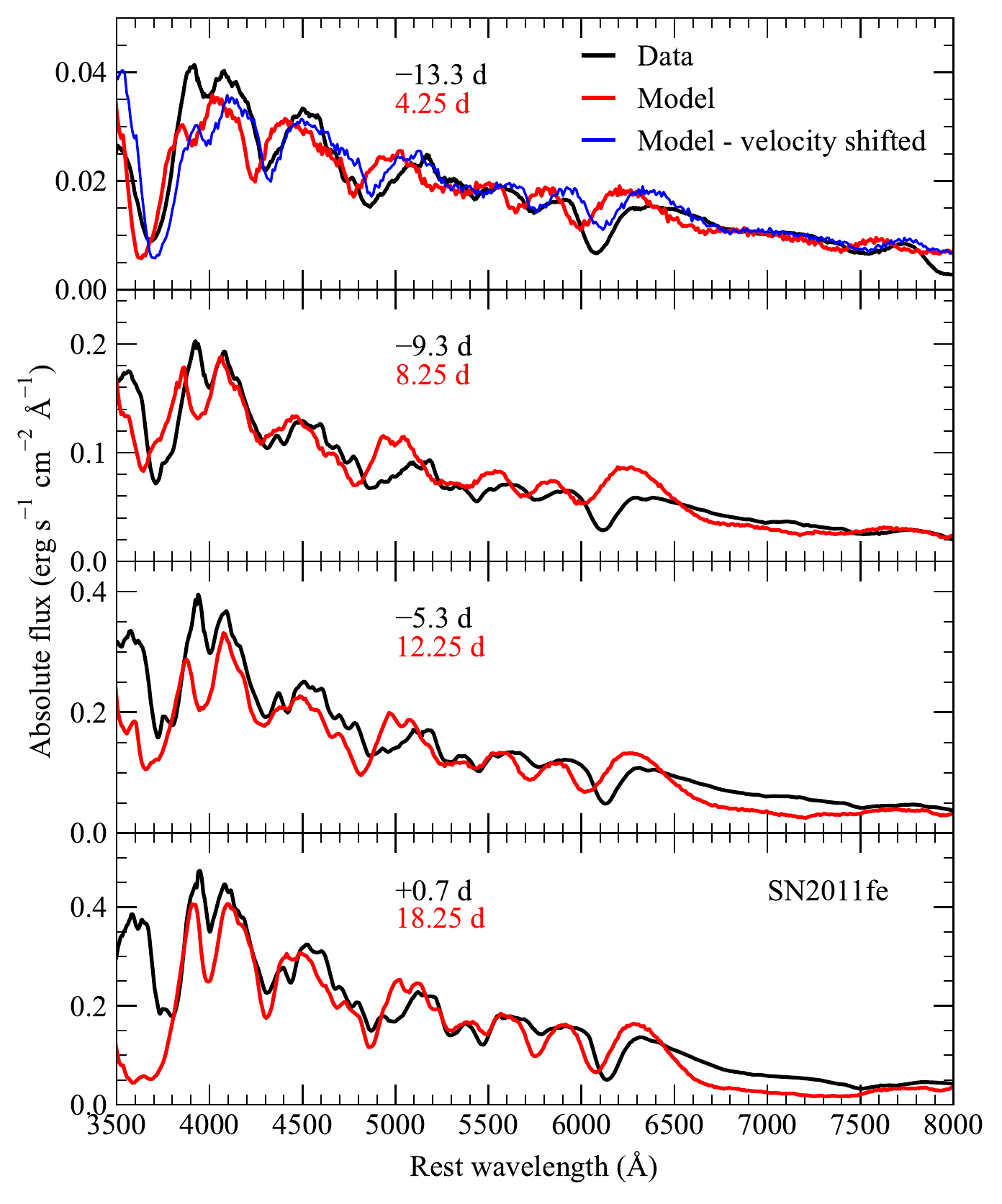}
    \end{subfigure}
    \begin{subfigure}[b]{0.48\textwidth}\includegraphics[width=\columnwidth]{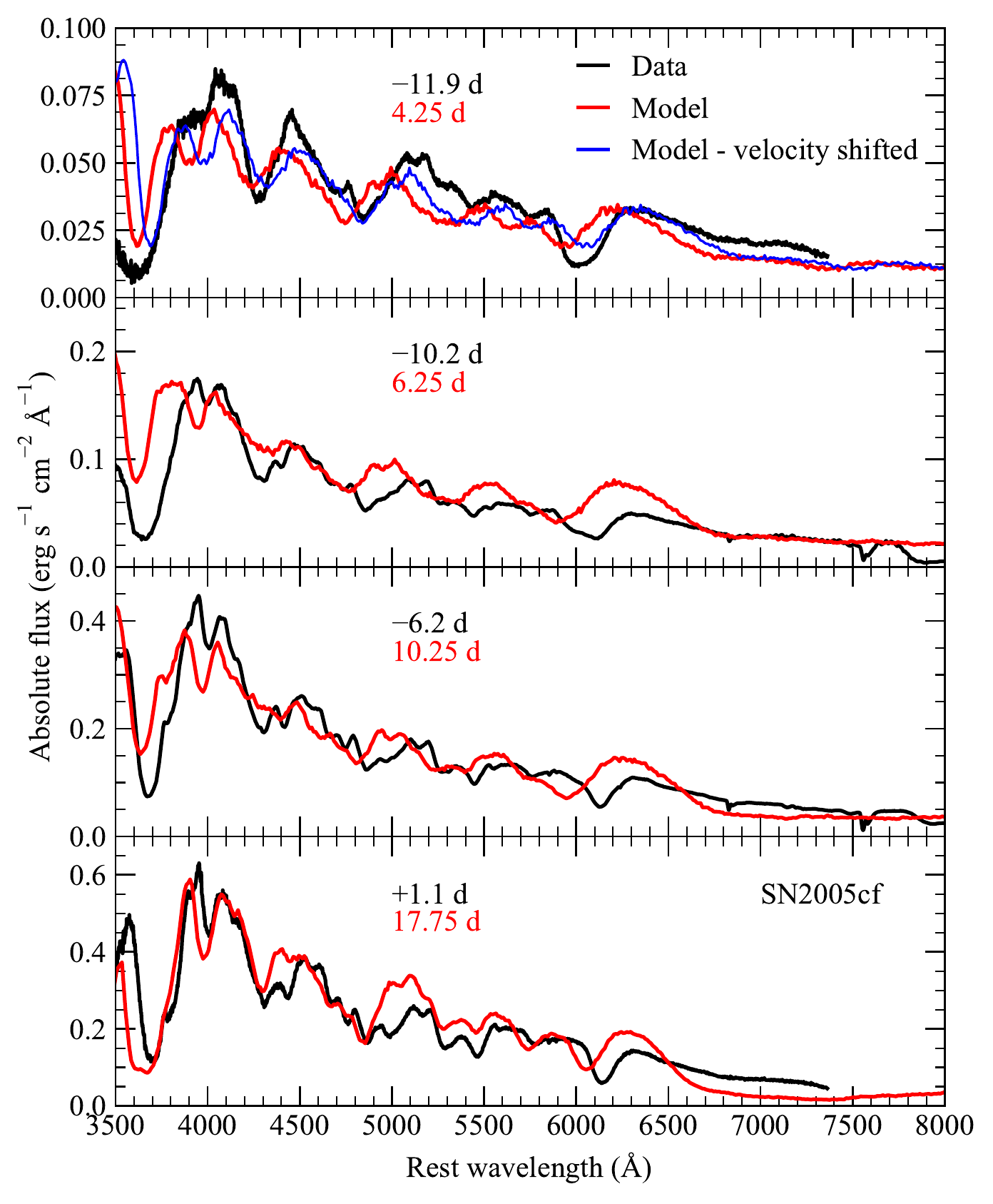}
    \end{subfigure} 
    \caption{Comparisons between spectra of  SNe~2011fe and 2005cf (black), and our double-detonation models with \sulphur{}-dominated shells (red) at epochs from 4 -- 18 d after explosion, along with the parameters of both models. Spectra are shown on an absolute flux scale. For the first epoch, we also show model spectra shifted in velocity to provide better agreement to the observed features (blue). The phases of the observed spectra of SNe~2011fe and 2005cf relative to $B$-band maximum are given in black, while the time since explosion for our model comparison spectra is shown in red.
    }
    \label{fig:normal_spec}
\end{figure*}

In the following section, we discuss whether our double detonation models are consistent with observations of SNe~Ia. We compare to light curves and spectra of two well-observed and prototypical SNe~Ia, SNe~2011fe \citep{11fe--nature, 2011fe, vinko--12} and 2005cf \citep{pastorello--07, garavini--07}. Figure~\ref{fig:normal_lc} shows the light curves of both objects compared to our models, while spectra are shown in Fig.~\ref{fig:normal_spec}. For SN~2011fe, we show a model with a 1.0~\mass{} core and 0.04~\mass{} shell dominated by sulphur (WD1.00\_He0.04\_BF0.20\_DP32S). The \nickel{} mass of this model (0.49~\mass{}) is comparable to estimates for SN~2011fe ($\sim$0.45~\mass{}; \citealt{11fe--nature}). For SN~2005cf, we find that a larger total mass is required to reproduce the higher core \nickel{} mass (0.7~\mass{}; \citealt{pastorello--07}). Our models with either a 1.0~\mass{} core and 0.10~\mass{} shell or 1.1~\mass{} core and 0.01~\mass{} shell both produce similar light curves and spectra for a \sulphur{}-dominated shell and may be considered interchangeable. Here we present the model with a 1.0~\mass{} core and 0.10~\mass{} shell  for SN~2005cf. As previously mentioned (Sect.~\ref{sect:construct_core_comp}), there is disagreement between various studies over the amount of \nickel{} produced for a given white dwarf mass during explosion. For this reason, the core and shell masses presented here should not be taken as predictions for the objects discussed here, but are simply given as reference to identify the comparison models shown. SN~2011fe has been corrected for a total extinction of $E(B-V)$ = 0.01~mag. \citep{11fe--nature}, while SN~2005cf has been corrected for $E(B-V)$ = 0.1~mag. \citep{pastorello--07}.

\par

Figure~\ref{fig:normal_lc} demonstrates that our double detonation models with \sulphur{}-dominated shells provide good agreement with the light curve shapes of both objects beginning a few days after explosion and extending to approximately maximum light. The largest discrepancies between models and observations are observed in the $U$-band, but we note this is likely related to the simplified composition and ejecta structure used (see \citealt{magee--20}), and will be explored in future work. \citet{townsley--19} have also previously shown that a double detonation model with a 1.0~\mass{} core and 0.02~\mass{} helium shell dominated by IMEs can reproduce the light curve of SN~2011fe. For epochs $\lesssim$4 days after explosion however, both models clearly show a rise that is too sharp and a dark phase that is too long to match the observed flux. By comparing to several Chandrasekhar mass models of different \nickel{} distributions, \citet{magee--20} found that the early light curve points of SN~2011fe can be reproduced by a relatively extended \nickel{} distribution with a mass fraction in the outer ejecta of $\sim$0.03. In contrast, the double detonation models shown here, which have \sulphur{}-dominated shells, do not contain any \nickel{} in the outer ejecta. As shown in Fig.~\ref{fig:compositions}, the functional form used for the models presented here produces a \nickel{} distribution that is somewhat more compact than that predicted by \citet{kromer--10}. A slightly more extended core \nickel{} distribution for the double detonation models with \sulphur{}-dominated shells could likely reproduce the earliest detections of SNe 2011fe and 2005cf, without adversely affecting the light curve at later times.

\par

Figure~\ref{fig:normal_spec} shows a comparison between the spectra of SNe~2011fe and 2005cf and their corresponding double detonation models with \sulphur{}-dominated shells. Previous comparisons to double detonation and bare sub-Chandrasekhar mass models have focused only on spectra around maximum light. Here we show spectra for both objects at multiple epochs, beginning $\sim$4 days after explosion and extending to maximum light. For our double detonation models, we find that spectra at 4.25\,d after explosion are consistent with those of both objects approximately two weeks before maximum. Although many of the features are reproduced with approximately the correct strength and shape, such as \ion{Si}{ii}~$\lambda$6\,355 \& 5\,972, \ion{S}{ii}-W, \ion{Mg}{ii}~$\lambda$4\,481, \ion{O}{i}~$\lambda$7\,774, and the \ion{Ca}{ii} NIR triplet, they are all noticeably offset to higher velocities in the models compared to the observed spectra. Therefore, in Fig.~\ref{fig:normal_spec}, we also show our model spectra with a velocity shift of $\sim$6\,000~km~s$^{-1}$ applied and find improved agreement. As discussed in Sect.~\ref{sect:construct_shell_comp}, the systematic shift to high velocities in our spectra could be due to simplifications made in our model density profiles, particularly in the outer regions. Closer to maximum light, velocities of many features (such as the \ion{S}{ii}-W feature) in our model spectra show good agreement with both SNe, although \ion{Si}{ii} velocities are still somewhat higher than those observed. In Sect.~\ref{sect:shell_composition}, we show how the material in the helium shell can impact the spectroscopic features within the first week of explosion. Qualitatively, this is similar to the broad \ion{Si}{ii}~$\lambda$6\,355 feature in SN~2005cf, which has been argued to have a high velocity component \citep{garavini--07}. Again we note that there is a systematic shift of all features to higher velocities at this time. Nevertheless, our models provide tentative evidence that the high velocity components in some SNe~Ia may be due to interactions with a helium shell containing IMEs.

\par

Our models verify the claims of \citet{kromer--10} and \citet{townsley--19} that double detonation explosions in which the helium shell does not produce significant fractions of IGE are consistent with the observed behaviour of normal SNe~Ia and therefore cannot be ruled out on this basis. Here, we extend this to show that models with different helium shell masses are also capable of reproducing multiple normal SNe~Ia and at various epochs up to maximum light. While our models generally show good agreement from a few days after explosion, the initial rise of the light curve is sharper than observed, supporting the claims of \citet{magee--20} that a more extended distribution for \nickel{} in the core may be required.

%

\section{Comparisons with SNe~Ia showing an early light curve bump}
\label{sect:comparisons_bump}

A handful of objects with early light curve bumps have now been discovered. Among these objects, two distinct groups are clearly apparent based on their optical colour close to maximum light -- those that are extremely red, with $B-V \gtrsim 1$ (SNe 2016jhr, 2018byg) and those that are relatively normal with a blue colour, $B-V \lesssim 1$ (SNe 2017cbv, 2018oh, 2019yvq, and iPTF14atg).
In Sect.~\ref{sect:10ops}, we discuss the case of PTF10ops separately, which has been suggested to show excess flux at early times \citep{jiang--18}, but its true nature is unclear due to a limited data set. In the following sections, we compare our models to observations of SNe~Ia with early bumps split into `blue' and `red' SN sub-classes, and discuss our models in the context of other scenarios that have been proposed for these objects. 

\par

Again, we stress that exact values for the core mass should not be taken literally due to the uncertainty in the amount of \nickel{} produced (Sect.~\ref{sect:construct_core_comp}), but are given for reference. The shell masses presented here are likely more robust predictions as our models cover a wide range of burned masses and products, and the observed shape of the bump will be highly sensitive to the mass of the shell.

\subsection{Blue SNe~Ia with an early bump}

\begin{figure*}
    \centering
    \begin{subfigure}[b]{0.49\textwidth}
        \includegraphics[width=8.6cm]{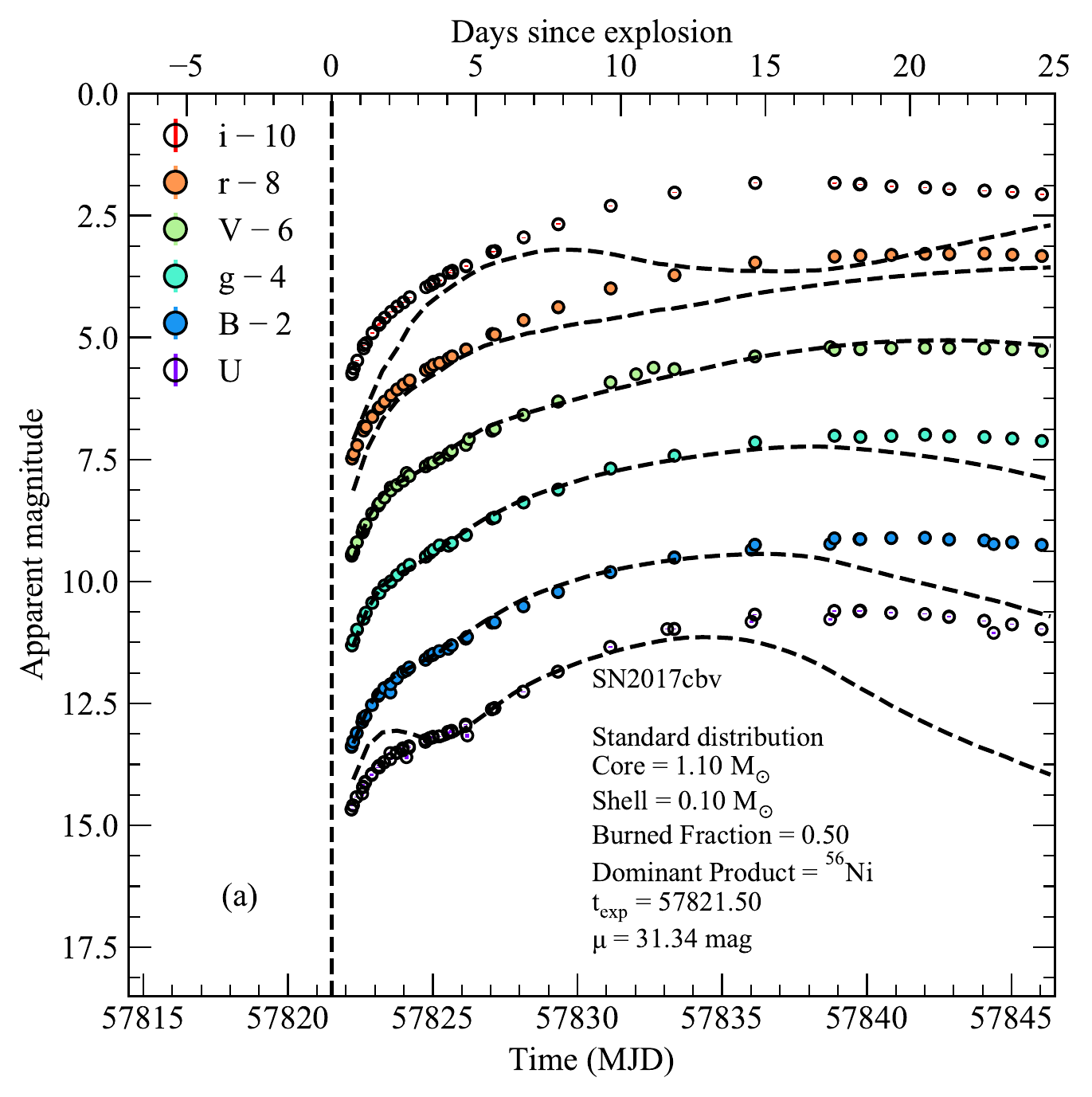}
    \end{subfigure}
    \begin{subfigure}[b]{0.49\textwidth}
        \includegraphics[width=8.6cm]{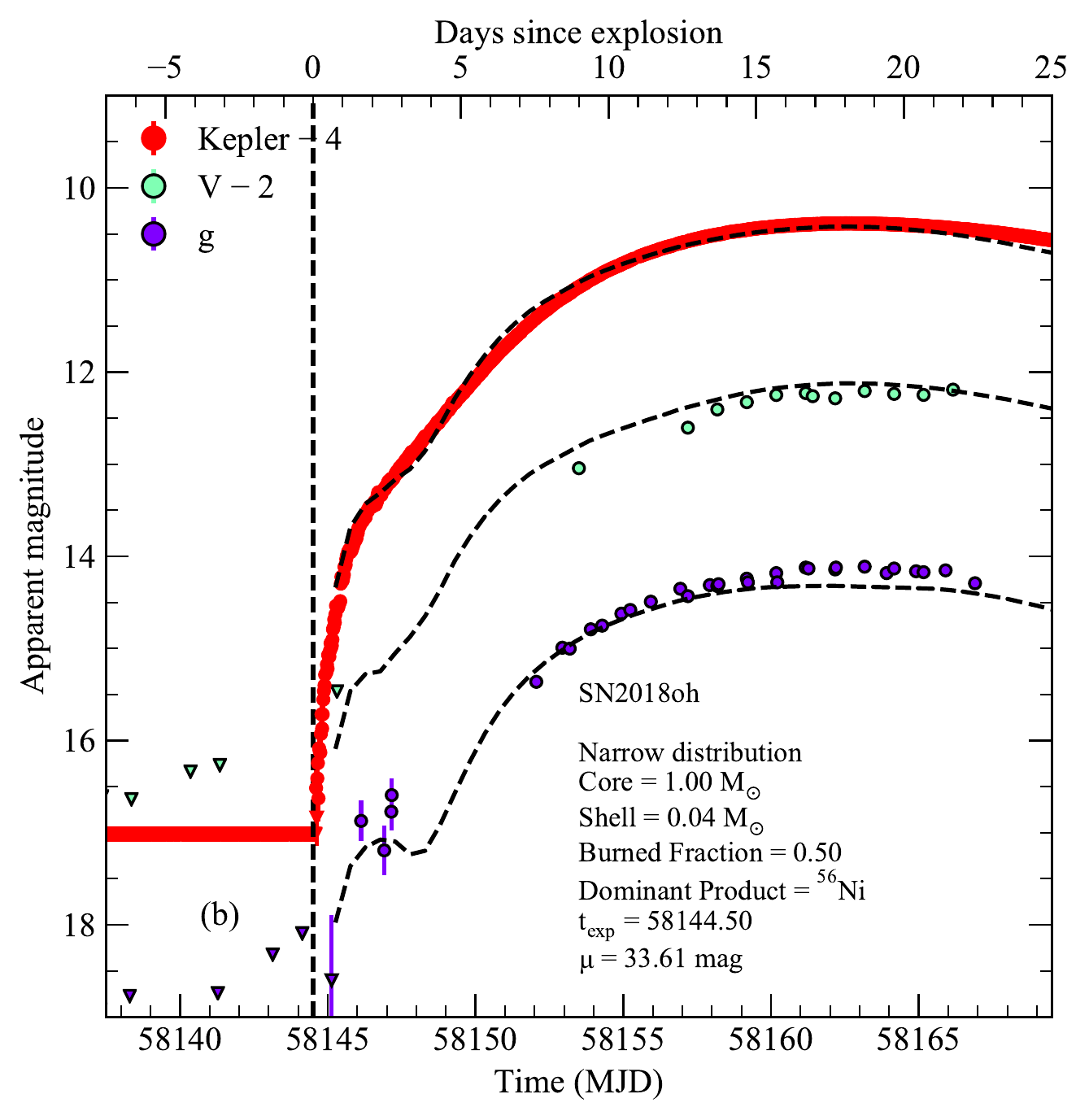}
    \end{subfigure} 
    \\
    \begin{subfigure}[b]{0.49\textwidth}
        \includegraphics[width=8.6cm]{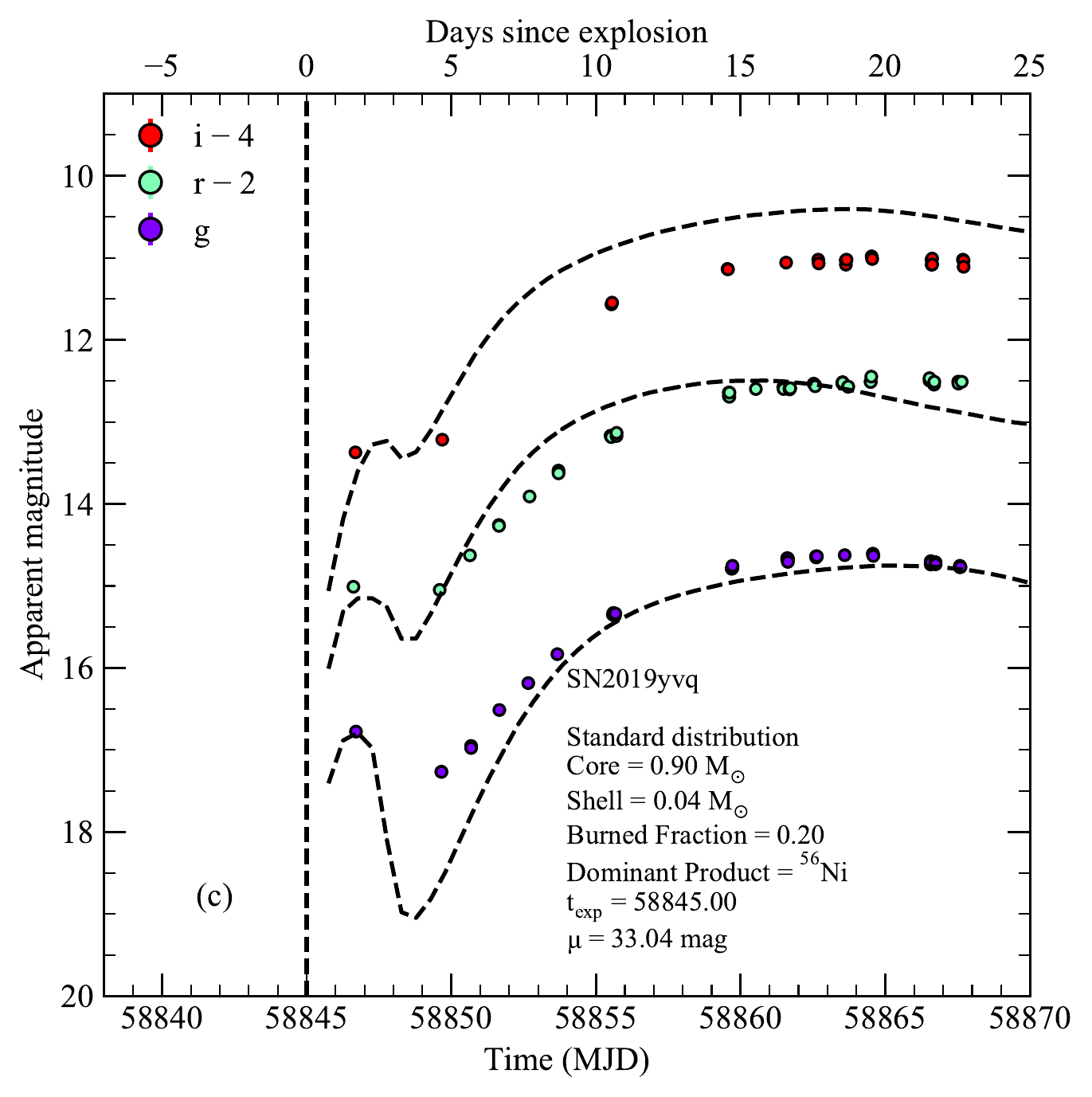}
    \end{subfigure}
    \begin{subfigure}[b]{0.49\textwidth}
        \includegraphics[width=8.6cm]{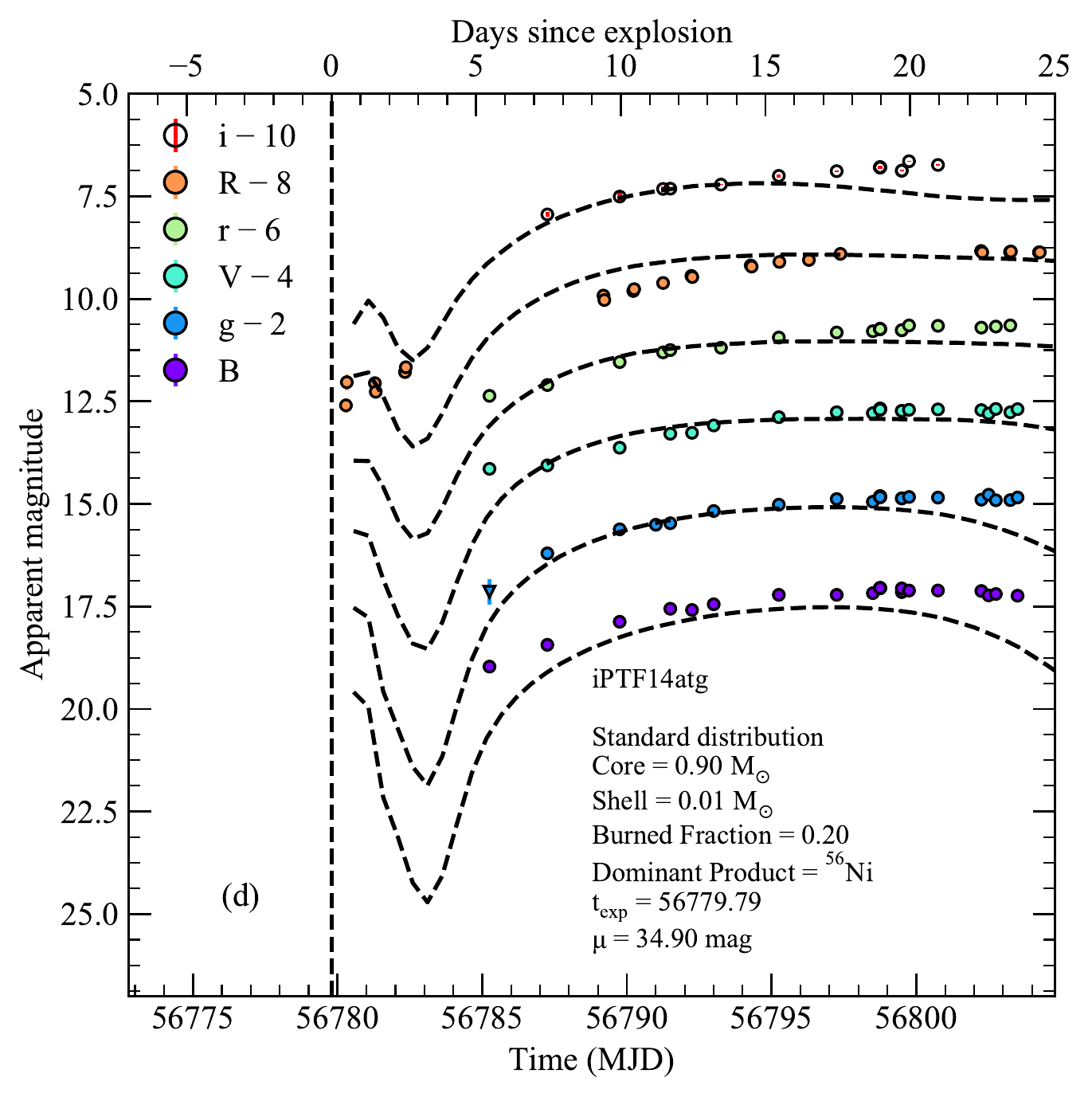}
    \end{subfigure}
    \caption{Comparisons between SNe~2017cbv, 2018oh, 2019yvq, and iPTF14atg (coloured circles) and our double detonation models (dashed lines) with \nickel{}-dominated shells. The model parameters and assumed distance modulus are given for each object. The estimated time of explosion (based on the comparison with the model light curve) is also shown as a vertical dashed line for each SN.
   }
    \label{fig:blue_bump_lc}
\end{figure*}

\begin{figure}
\centering
\includegraphics[width=\columnwidth]{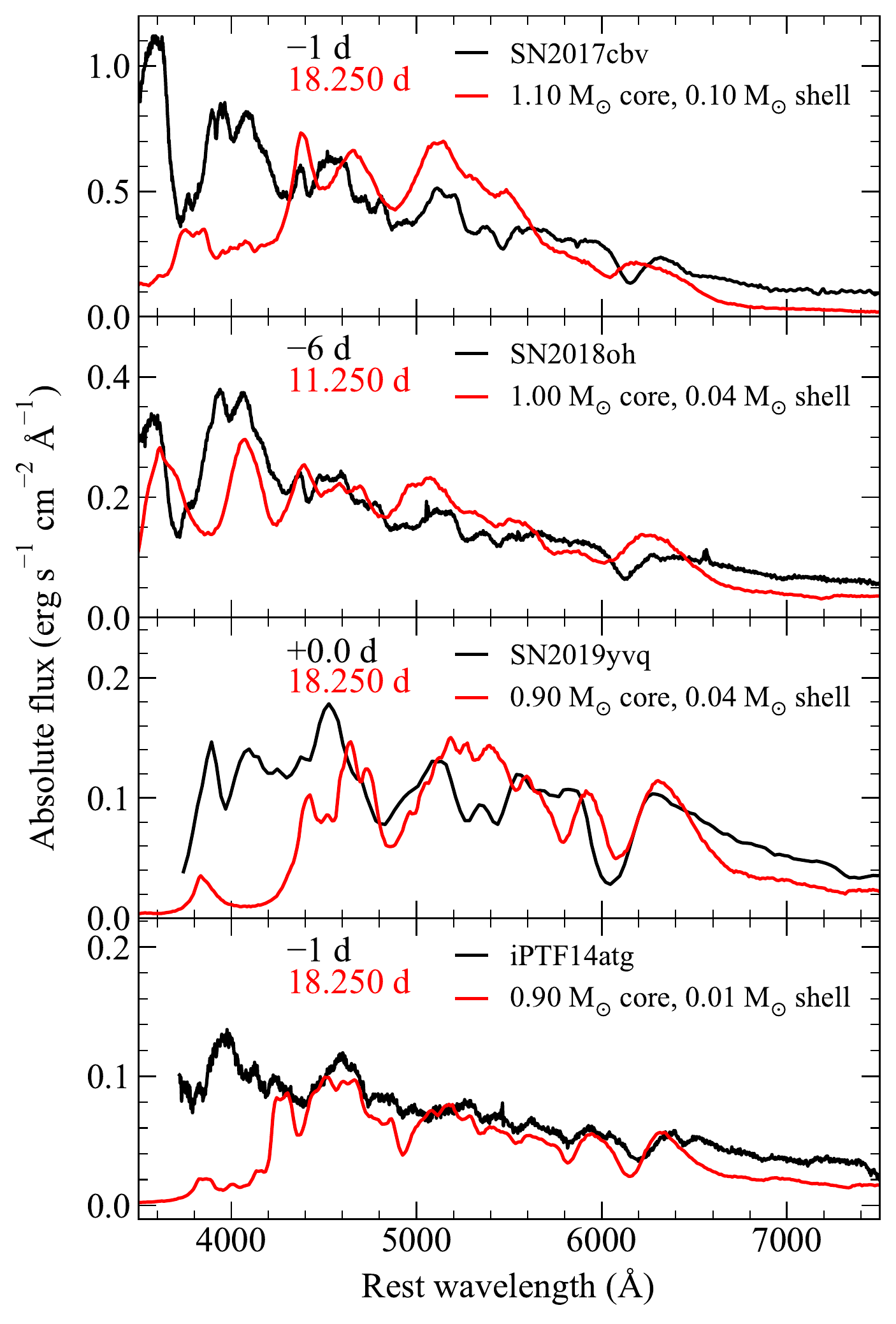}
\caption{Comparisons between SNe~2017cbv, 2018oh, 2019yvq, and iPTF14atg (black) and our model spectra (red) around maximum light, along with the parameters of all models. Spectra are shown on an absolute flux scale. Phases of SNe relative to $B$-band maximum are given in black, while the time since explosion for our model spectra is shown in red. }
\label{fig:blue_bump_spec}
\centering
\end{figure}

Figure~\ref{fig:blue_bump_lc} shows a comparison between the light curves of SNe~2017cbv, 2018oh, 2019yvq, iPTF14atg, and four of our double detonation models with \nickel{}-dominated shells. These models are broadly able to reproduce the shape of the early light curve bump. In Fig.~\ref{fig:blue_bump_spec}, we compare spectra of each SN and model around maximum light.

\par

The light curve of SN~2017cbv was previously compared to double detonation models by \citet{maeda--18}, however their models were somewhat too faint (SN~2017cbv was a bright SN~Ia and showed a peak absolute magnitude of $M_{\rm{B}}$ = $-20.4$; \citealt{hosseinzadeh--17}). In Fig.~\ref{fig:blue_bump_lc}(a), we show that our model with a 1.1~\mass{} core and massive helium shell of 0.10~\mass{} dominated by \nickel{} can generally reproduce the shape of the early light curve of SN~2017cbv. The \nickel{} mass of this model is 0.94~\mass{}. For  the model shown in Fig.~\ref{fig:blue_bump_lc}(a), the $U$-band shows a more pronounced bump than observed; we speculate that minor changes to the composition within the helium shell could be made to find improved agreement in this band. Regardless of this discrepancy, beginning approximately three weeks after explosion the model light curves show a much faster decline in the bluer bands ($U$, $B$, and $g$) than SN~2017cbv. This is further demonstrated by Fig.~\ref{fig:blue_bump_spec}, which shows that the maximum light spectrum of our double detonation model with a \nickel{}-dominated shell exhibits significant line blanketing that is inconsistent with SN~2017cbv. In addition, the spectral features are also inconsistent with SN~2017cbv. Our model shows a significantly broadened \ion{Si}{ii}~$\lambda$6\,355 feature that has blended with \ion{Si}{ii}~$\lambda$5\,972. In contrast, the observed spectra of SN~2017cbv show a well defined \ion{Si}{ii}~$\lambda$6\,355 feature and lacks \ion{Si}{ii}~$\lambda$5\,972.

\par

\citet{hosseinzadeh--17} compare observations of SN~2017cbv to models of the interaction between the SN ejecta and a non-degenerate companion star. While they find that interaction with a 56~$R_{\rm{\odot}}$ sub-giant is able to reproduce the bump in the optical bands, the model over-predicts the flux in the UV bands ($UVW1$, $UVM2$, and $UVW2$). \citet{hosseinzadeh--17} argue that this could indicate the flux resulting from the interaction is not a black-body, or an alternative explanation is required. Based on nebular spectra hundreds of days after explosion, \citet{sand--18} rule out the presence of any significant H$\alpha$ signatures, which are expected if material is stripped from a non-degenerate companion. Finally, the presence of CSM was also ruled out by \citet{ferretti--17}. Our models indicate that SN~2017cbv likely did not result from a double detonation explosion. Taken together, the nature of SN~2017cbv is still uncertain.

\citet{dimitriadis--19} compare SN~2018oh to a double detonation explosion model with a 0.98~\mass{} core and 0.05~\mass{} helium shell, which produces 0.45~\mass{} of \nickel{}. To match the early light curve bump of SN~2018oh, \citet{dimitriadis--19} invoke mixing of the SN ejecta after explosion. Therefore, the model does not produce a well defined early bump in the light curve, but rather shows an extended `shoulder' that more closely resembles SN~2018oh. It is not clear however, whether such mixing could be achieved in double detonation explosions. Figure~\ref{fig:blue_bump_lc}(b) shows that our model with a 1.0~\mass{} core and thin helium shell of 0.04~\mass{} (i.e. a \nickel{} mass of 0.49~\mass{})  with a narrow isotope distribution dominated by \nickel{} provides reasonable agreement to the light curve shape of SN~2018oh. A narrow isotope distribution is required to reduce the mass of \iron{} in the shell and avoid the peak produced by its short lifetime. Even for this distribution however, the bump in the model Kepler band light curve is more pronounced than observed in SN~2018oh. Reducing the \iron{} mass further may again provide improved agreement as \citet{magee--20b} have shown that the light curve bump in SN~2018oh can be reproduced by a model containing a clump of pure \nickel{} in the outer ejecta.

\par

Although this model is less affected by line blanketing than the model for SN~2017cbv, due to the lower mass of the helium shell, Fig.~\ref{fig:blue_bump_spec} shows that the spectral features of SN~2018oh are also inconsistent with a double-detonation scenario. \citet{dimitriadis--19} did not consider the spectra of their double-detonation models compared to SN~2018oh. In summary, our models indicate that SN~2018oh did not result from a double-detonation explosion as we are unable to simultaneously match the spectroscopic features and early light curve bump, regardless of the composition of the helium shell.

\par

The favoured interpretation for SN~2018oh by \citet{dimitriadis--19} is that of interaction with a non-degenerate companion. Again similar to SN~2017cbv, no evidence for material stripped from a non-degenerate companion has been found in late-time spectra of SN~2018oh \citep{tucker--19}. The case of interaction with CSM was investigated by \citet{shappee--2019}, who found that none of their interaction models could satisfactorily reproduce the initial early light curve shape of SN~2018oh. An alternative case of interaction for SN~2018oh was suggested by \citet{levanon--19}. In this scenario, the explosion follows from the merger of two white dwarfs. An accretion disk forms around the primary \citep{raskin--12, zhu--13}, which serves as the source of the CSM. After explosion, the SN ejecta shocks material in the disk, producing a flash of UV radiation that may be similar to that of SN~2018oh \citep{levanon--15}.  This scenario warrants further investigation with radiative transfer simulations, as alternatives appear to be ruled out for SN~2018oh.

\par

In Fig.~\ref{fig:blue_bump_lc}(c), we compare a double-detonation model to SN~2019yvq \citep{miller--20}. SN~2019yvq was a somewhat peculiar SN -- it was slightly under-luminous, but showed high velocity spectral features. \citet{miller--20} compared observations of SN~2019yvq to a variety of models, including double detonation explosions. They found reasonable agreement to a double-detonation model with a 0.92~\mass{} core and 0.04~\mass{} shell. In Fig.~\ref{fig:blue_bump_lc}(c) we show our model with comparable values -- a 0.9~\mass{} core and 0.04~\mass{} shell dominated by \nickel{}. The similarity of these values is unsurprising, given that \citet{miller--20} use the same modelling treatment as \citet{polin--19}, upon which our models are at least partially based. As in \citet{miller--20}, our model generally matches the early light curve bump in the redder bands ($r$ and $i$), but the $g$-band shows a larger decrease in magnitude than is observed immediately following the peak of the bump. Even for models with different dominant products (e.g. \iron{} and \chromium{}) in the shell, we are not able to simultaneously match the bump in both the $g$- and $r$-bands. As with SNe~2017cbv and 2018oh, Fig.~\ref{fig:blue_bump_spec} shows that the maximum light spectrum of SN~2019yvq is significantly bluer than the model and does not exhibit strong line blanketing. 

\par

In addition to double detonation explosions, \citet{miller--20} investigate other scenarios to explain SN~2019yvq, including an excess of $^{56}$Ni in the outer ejecta, a violent merger, and companion interaction. None of the proposed scenarios fully explain all of the observed features. One possible exception is the violent merger scenario. While it is likely that this scenario does produce some CSM, models including this material are currently unavailable. Based on the identification of calcium in nebular spectra and favourable comparisons with model nebular spectra, \citet{siebert--20} argue that SN~2019yvq was indeed the result of a helium shell detonation. The light curve curve of the model favoured by \citet{siebert--20} however, does not reproduce what is observed in SN~2019yvq. The model is simultaneously too bright and does not show a pronounced bump at early times. Whether it is possible to simultaneously match the early- and late-time observations of SN~2019yvq requires further investigation.  

\par

Finally, Fig.~\ref{fig:blue_bump_lc}(d) shows iPTF14atg compared to a double-detonation model with a 0.9~\mass{} core and thin helium shell of 0.01~\mass{}. This model contains only 0.13~\mass{} of \nickel{}. Unlike the other objects discussed here, the early bump observed in iPTF14atg was less pronounced in the optical bands, but clearly apparent at UV wavelengths. \citep{cao--15}. The origin of this early excess was discussed by \citet{cao--15}, who argued that it is consistent with theoretical predictions of the collision between the SN ejecta and companion star. They also discuss the possibility of this excess arising from \nickel{} at the surface of the ejecta, such as in double detonation models, and estimate that this would require a \nickel{} mass of $\sim$0.01~\mass{} at the surface. Figure~\ref{fig:blue_bump_lc}(d) shows that even for our model with a 0.01~\mass{} shell, the early light curve bump produced is inconsistent with iPTF14atg. Figure~\ref{fig:blue_bump_spec} also shows the maximum light spectrum of iPTF14atg. Our model predicts a spectrum at maximum light that is substantially redder than iPTF14atg, and shows strong flux suppression for wavelengths $\lesssim$4\,200~\AA. The origin of iPTF14atg was also considered by \citet{kromer--16}, who favour the violent merger of two white dwarfs. This particular realisation of the violent merger scenario did not predict a light curve bump, but interaction due to the presence of some CSM for similar models may be consistent with the observations.

\subsection{Red SNe~Ia with an early bump}

\begin{figure}
\centering
\includegraphics[width=\columnwidth]{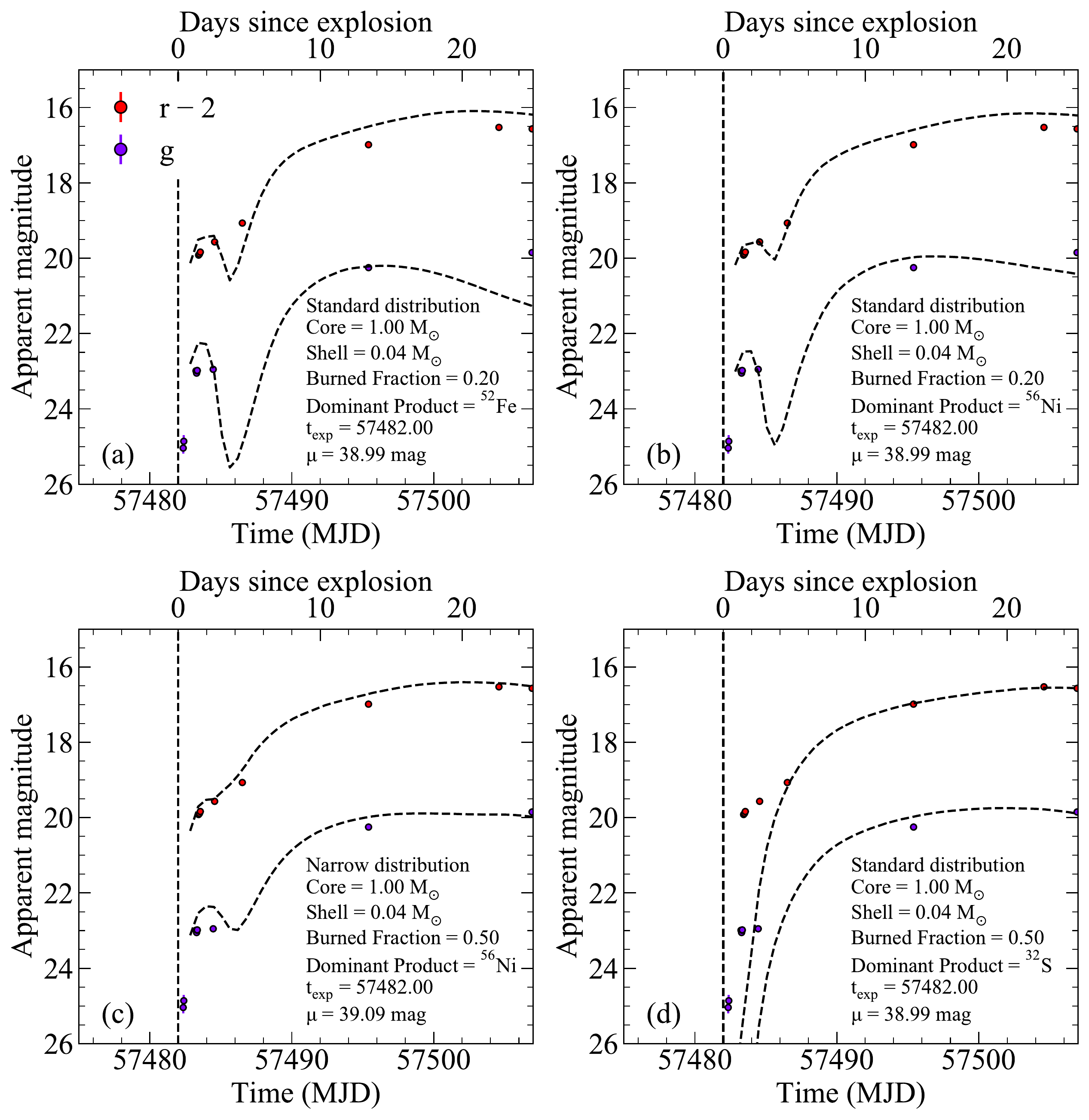}
\caption{Comparisons between SN~2016jhr and a subset of our models. The model parameters and assumed distance modulus are given for each object. The estimated time of explosion (based on agreement with the model light curve) is also shown as a vertical dashed line for each SN. In all cases, our model light curves have been transformed into the observer frame of SN~2016jhr (redshift of 0.117).
}
\label{fig:16jhr_lc}
\centering
\end{figure}

Observations of SN~2016jhr were presented by \citet{jiang--2017}, who found reasonable agreement with models of double detonations and either sub- or near-Chandrasekhar mass white dwarfs. In Fig.~\ref{fig:16jhr_lc}, we show a comparison between some of our double-detonation models and SN~2016jhr. We note that unlike \citet{jiang--2017}, we do not apply K-corrections to the observed light curve as these may be uncertain due to the peculiar nature of SN~2016jhr. Instead, we transform our model spectra into the observer frame (redshift of 0.117) and calculate observer frame light curves. For all comparisons, we assume an explosion date of MJD = 57482.0, which is a few hours before the first detection. We also assume a distance modulus of $\mu$ = 38.99~mag, which is 0.3~mag higher than that used by \citet{jiang--2017}.

 Figure~\ref{fig:16jhr_lc}(a) shows our model with a 1.0~\mass{} core and 0.04~\mass{} helium shell dominated by \iron{}. The core and shell mass of models shown here is comparable to the models presented by \citet{jiang--2017} (1.03~\mass{} core and 0.054~\mass{} helium shell) and \citet{polin--19} (1.0~\mass{} core and 0.05~\mass{} helium shell). Our model contains 0.49~\mass{} of \nickel{} and is able to broadly reproduce the light curve shape of SN~2016jhr during the first few days after explosion, but shows a much faster decline in the $g$-band than observed. Assuming a shell dominated by \nickel{} (Fig.~\ref{fig:16jhr_lc}(b)), we again find that our model can reproduce the early light curve bump. We also find improved agreement in the $g$-band close to maximum light, however the model still declines somewhat faster than observed. In Fig.~\ref{fig:16jhr_lc}(c), we show a model assuming our narrow isotope distribution. In this model, a much larger fraction of \nickel{} is present in the shell relative to \iron{} than in our standard distribution. In this case, the $r$-band light curve does not display a pronounced bump and instead shows a shoulder to the light curve that is still generally consistent with SN~2016jhr. For the $g$-band, this model is clearly brighter during the bump than the observations, however a lower burned fraction may produce more favourable agreement (the models shown in Fig.~\ref{fig:16jhr_lc}(a)~\&~(b) both have burned fractions of 0.2, while the model in Fig.~\ref{fig:16jhr_lc}(c) has a burned fraction of 0.5). Around maximum light, this model also produces a broader $g$-band light curve that more closely resembles SN~2016jhr. As a further point of comparison, we also show a model with a \sulphur{}-dominated shell (Fig.~\ref{fig:16jhr_lc}(d)). In this case, the model clearly does not reproduce the early light curve bump, but provides good agreement around maximum light.

\begin{figure}
\centering
\includegraphics[width=\columnwidth]{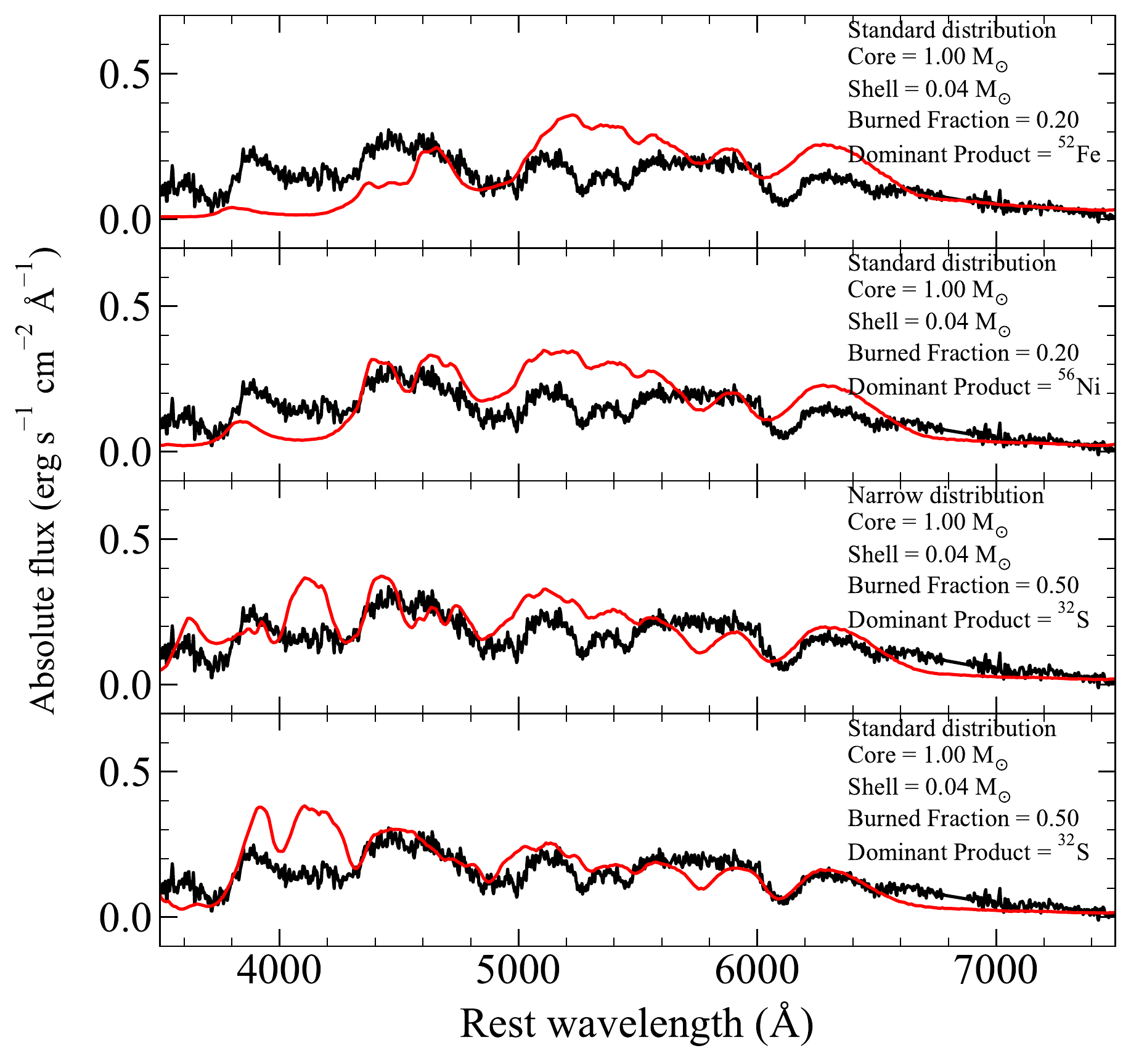}
\caption{Comparisons between SN~2016jhr (black) and our model spectra (red) around maximum light, along with the parameters of both models.
}
\label{fig:16jhr_spec}
\centering
\end{figure}

In Fig.~\ref{fig:16jhr_spec}, we show spectra for the models presented in Fig.~\ref{fig:16jhr_lc} at 18.25\,d after explosion and compare them to SN~2016jhr approximately two days before maximum light. As expected from the light curve comparison, Fig.~\ref{fig:16jhr_spec} shows that our standard isotope distribution models with \iron{}- and \nickel{}-dominated shells do a reasonable job of reproducing the maximum light spectrum. The \nickel{}-dominated shell produces better agreement at shorter wavelengths, while the \iron{}-dominated shell model shows more extreme flux suppression. In both cases, the continuum flux around $\sim$5\,000 -- 5\,550~\AA\, is higher than observed. The narrow distribution model shows a \ion{Si}{ii}~$\lambda$6\,355 velocity closer to SN~2016jhr, but also does not manage to reproduce the spectrum at shorter wavelengths. Again, we speculate that a smaller mass of burned material would produce improved agreement in this case. For our \sulphur{}-dominated shell model, we find that the model spectrum provides excellent agreement with the velocities of IMEs, such as \ion{Si}{ii} and \ion{S}{ii}. In contrast to the other models shown in Fig.~\ref{fig:16jhr_spec}, the \sulphur{}-dominated model does not show enough flux suppression at shorter wavelengths and instead is bluer than SN~2016jhr. 

\par

Taken together, our models corroborate the claims of \citet{jiang--2017} that SN~2016jhr is consistent with a double detonation explosion. The exact composition of the shell required is unclear, although it must include at least some amount of short-lived radioactive isotopes. Minor changes to the models presented here could provide improved agreement. Assuming a helium shell dominated by IMEs, we also find good agreement with the light curves and spectrum close to maximum light, although the model spectrum is too blue, which could indicate an alternative explanation for the early light cure bump is also possible. Indeed, a small \nickel{} excess in the outer ejecta may also provide good agreement with the early light curve shape and produce a redder spectrum consistent with SN~2016jhr.

\begin{figure}
\centering
\includegraphics[width=\columnwidth]{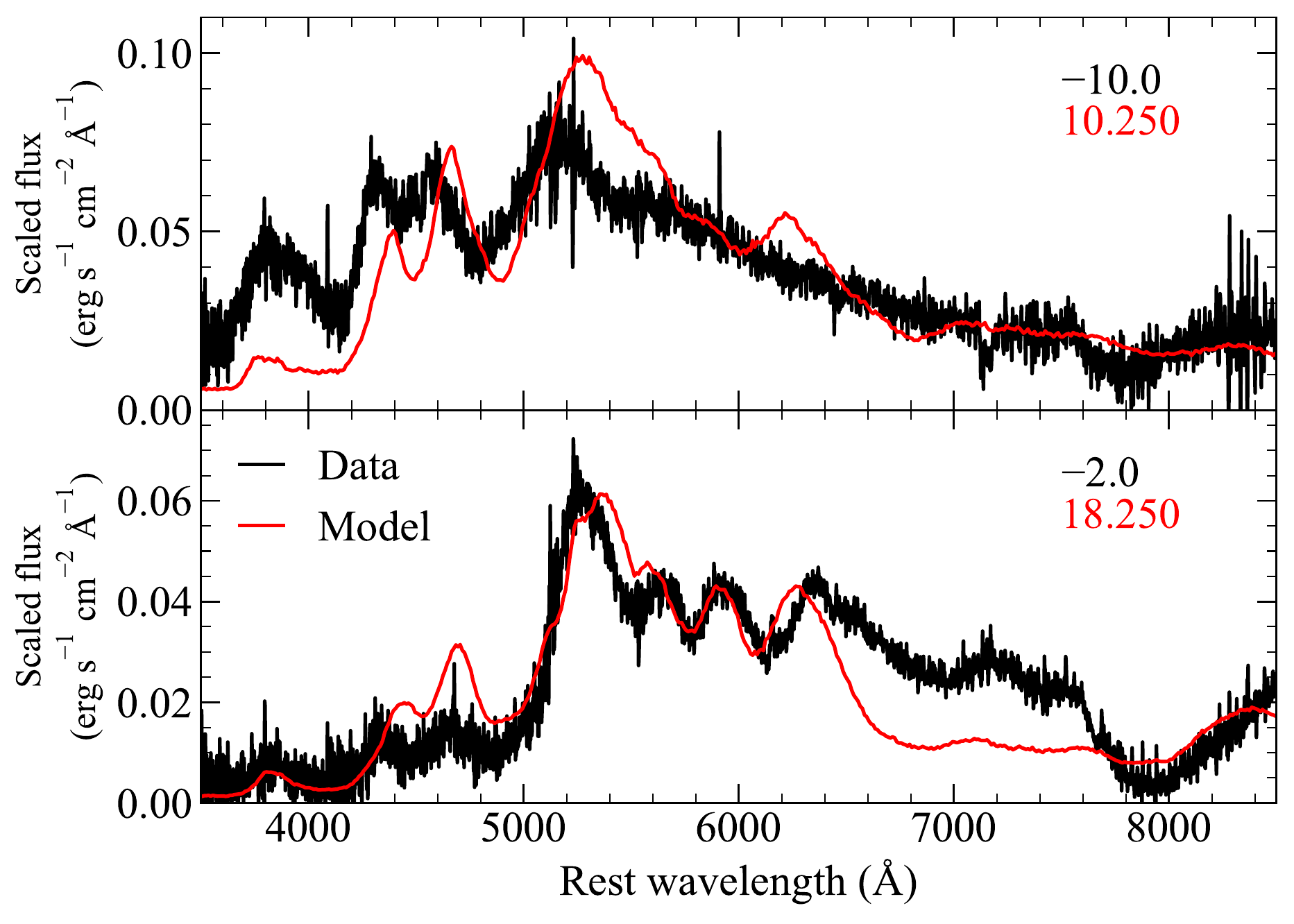}
\caption{Comparison between spectra of SN~2018byg (black) and our model with a 0.9~\mass{} core and 0.1~\mass{} shell (red). Phases of SN~2018ybq are given relative to $r$-band maximum, while days since explosion are given for our model. Spectra are shown in scaled flux.}
\label{fig:18byg_spec}
\centering
\end{figure}

In addition to SN~2016jhr, SN~2018byg also shows a peculiar early light curve and extremely red colours close to maximum light. The observations of SN~2018byg were presented by \citet{de--19}, who show that SN~2018byg displays a shoulder to the early rise of the $r$-band light curve. \citet{de--19} argue that SN~2018byg is consistent with the double detonation of a low mass white dwarf core ($\sim$0.75~\mass{}) and massive helium shell ($\sim$0.15~\mass{}). \citet{de--19} were unable to reproduce the early light curve shape of SN~2018byg within the standard double detonation scenario and find all models produce a significant light curve bump that is not observed. Instead, \citet{de--19} artificially performed mixing of the ejecta to match the light curve shape, as in \citet{dimitriadis--19} for the case of SN~2018oh. Again, it is not clear how such mixing could be achieved.

\par

As the parameter space of our model set does not cover the appropriate \nickel{} mass range predicted for SN~2018byg, our models are all much brighter than the observations. Therefore, in Fig.~\ref{fig:18byg_spec}, we show spectra for one of our models with a low mass white dwarf (0.9~\mass{}) and thick helium shell (0.1~\mass{}) dominated by \nickel{} that is scaled to match the flux of SN~2018byg. Figure~\ref{fig:18byg_spec} shows that approximately 10\,d after explosion, our model generally reproduces the spectrum of SN~2018byg at $-$10\,d relative to maximum light. SN~2018byg shows a relatively flat continuum between $\sim$5\,500 -- 7\,000~\AA, while our model shows high-velocity \ion{Si}{ii}. Closer to maximum light, our model provides excellent agreement with SN~2018byg and is able to reproduce the extreme flux suppression at wavelengths $\lesssim$5\,000~\AA\, as well as the \ion{Si}{ii} features around $\sim$6\,000\,\AA.

\subsection{PTF10ops}
\label{sect:10ops}

PTF10ops was a peculiar SN~Ia that showed a light curve significantly broader than expected for its low luminosity ($M_{\rm{B}} = -17.77\pm0.04$; \citealt{maguire--11}). The early light curve of PTF10ops is not well sampled and therefore it is unclear whether it belongs to the group of SNe showing bumps at early times. For this reason, we opt to discuss it separately. \citet{jiang--18} argued that there was evidence for an early flux excess, however they stress this is based on a single point. First detection occurred approximately 17\,d before $B$-band maximum, while the next detection was six days later. Making a definitive statement on the origin of PTF10ops is therefore a challenging prospect. Here we discuss whether PTF10ops is consistent with models of double detonation explosions. 

\begin{figure}
\centering
\includegraphics[width=\columnwidth]{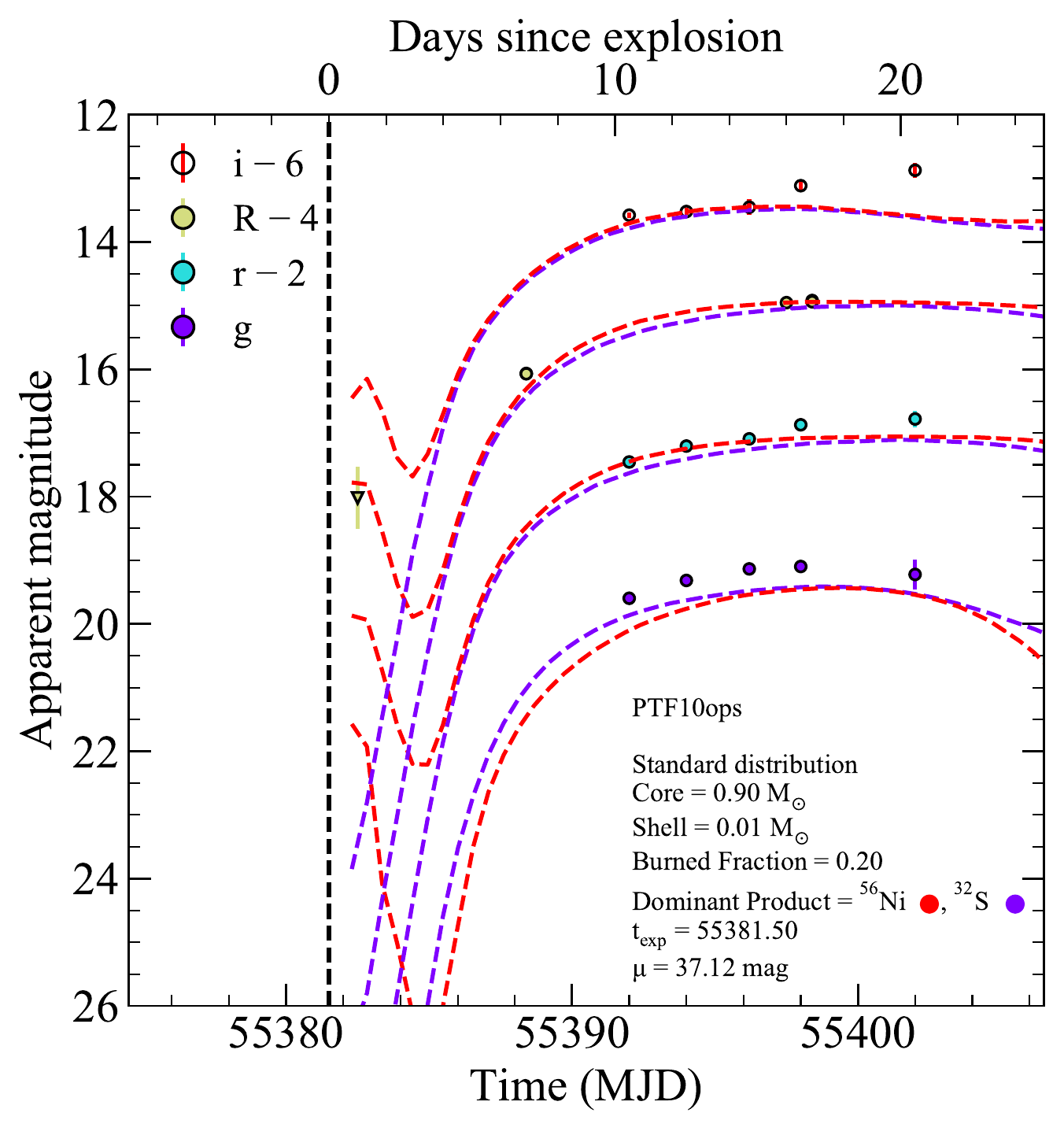}
\caption{Comparison between the light curve of PTF10ops and our models with a 0.9~\mass{} core, 0.01~\mass{} shell and either a \nickel{}-dominated (red) or \sulphur{}-dominated (purple) composition for the shell. The explosion epoch is shown as a vertical dashed line. 
}
\label{fig:10ops_lc}
\centering
\end{figure}

\par

In Fig.~\ref{fig:10ops_lc}, we show a comparison between the light curve of PTF10ops and models with either a \nickel{}- or \sulphur{}-dominated shell. These models contain a core mass of 0.9~\mass{} and shell mass of 0.01~\mass{}. The mass of \nickel{} produced in the core is 0.13~\mass{}. For the \nickel{}-dominated shell, our model shows a short-lived bump approximately one day after explosion that is consistent with the earliest detection of PTF10ops. The lack of detections in the following days and in other bands means that the decline from this initial bump could have simply been missed. This model also provides a good match to the later light curve evolution, but is slightly too faint in the $g$-band. Conversely, the \sulphur{}-dominated model does not match the earliest detection. Again, this model is able to match the light curve towards maximum light, but remains too faint in the $g$-band.

\begin{figure}
\centering
\includegraphics[width=\columnwidth]{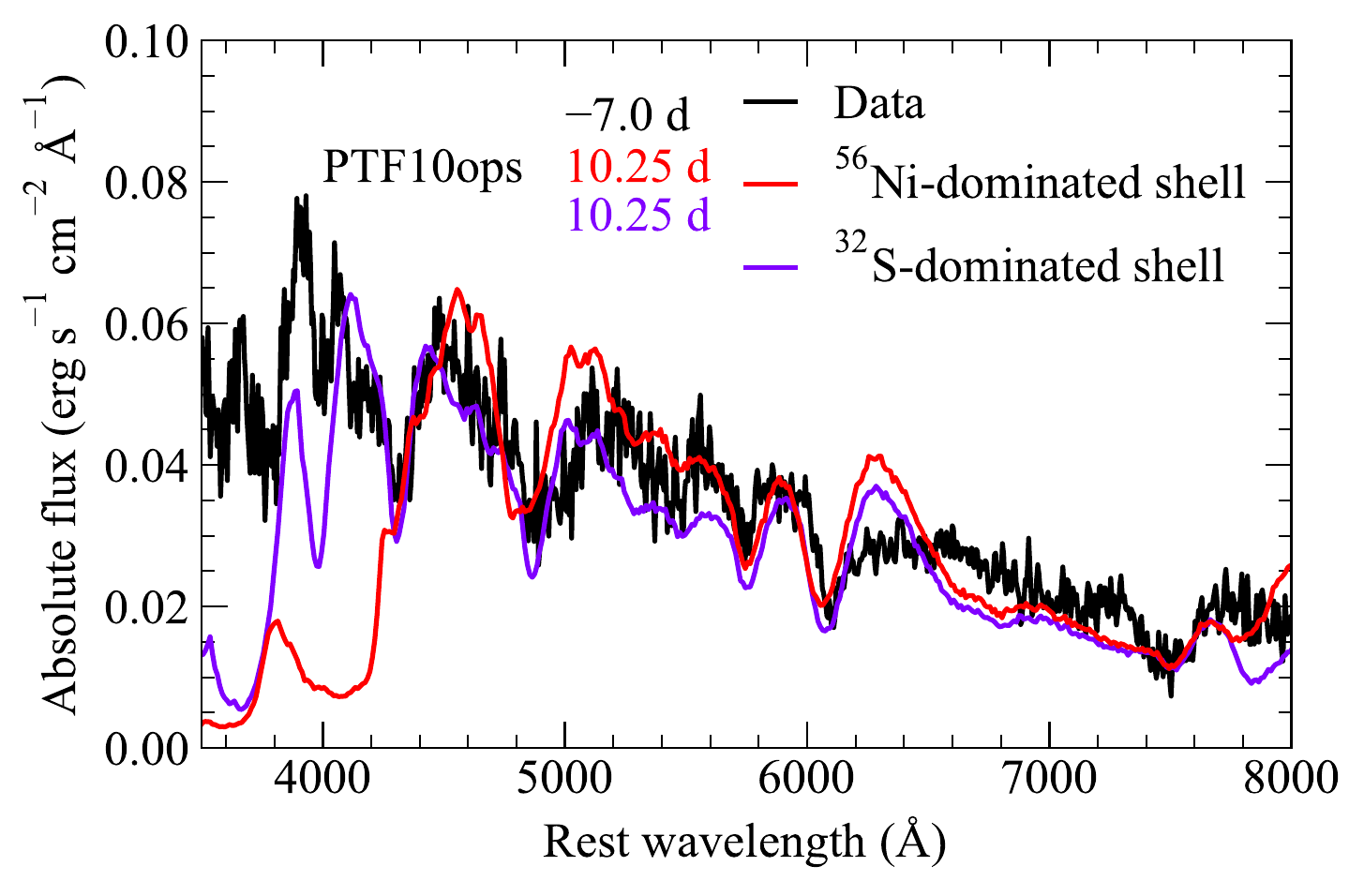}
\caption{Comparison between the spectrum of PTF10ops approximately one week before $B$-band maximum (black) and our models with a 0.9~\mass{} core, 0.01~\mass{} shell and either a \nickel{}-dominated (red) or \sulphur{}-dominated (purple) composition for the shell. Phases for the model spectra are given as days since explosion. All spectra are shown in an absolute flux scale. 
}
\label{fig:10ops_spec}
\centering
\end{figure}
    
Figure~\ref{fig:10ops_spec} shows spectra for both models compared to PTF10ops approximately one week before maximum light. While both models provide good agreement for wavelengths $\gtrsim$4\,500~\AA, it is clear that the \nickel{}-dominated shell model shows significant line blanketing that is inconsistent with PTF10ops. The \sulphur{}-dominated shell model provides improved agreement, although it shows much stronger \ion{Ca}{ii} features than those observed.

\par

Our models show that if PTF10ops did indeed have an excess of flux at early times, this was not due to a helium shell detonation as its spectra do not show significant line blanketing. Double detonation models in which the helium shell is dominated by IMEs provide better agreement overall, but they are not able to match the first detection in the light curve. As with normal SNe~Ia (Sect.~\ref{sect:comparisons_normal}), this could indicate that a somewhat extended \nickel{} distribution may be required for these sub-Chandrasekhar mass models. Alternatively, any early excess emission could be due to interaction, but the lack of nebular spectra and indeed the poorly sampled early light curve makes a definitive conclusion about the nature of PTF10ops a challenging prospect.

\subsection{Summary}
By comparing our models to observations of SNe~Ia with early bumps, we show that a variety of shell masses and compositions is necessary to reproduce the diversity observed. While double detonation models can match the shapes of the early bumps for all objects (with the exception of iPTF14atg), only those with red colours at maximum light are well matched both photometrically and spectroscopically throughout their evolution following the bump. An investigation of the extent to which double detonations can explain these blue objects requires further observations for a larger sample of objects. We note however, that all of the currently proposed mechanisms for producing early light curve bumps appear inconsistent with these blue objects in at least some way. These discrepancies may be due to incorrect colours or lacking features predicted from models of companion and CSM interaction in nebular spectra. 


%

\section{Conclusions}
\label{sect:conclusions}

We have presented a large-scale parameter study of the double detonation explosion scenario. Using the Monte Carlo radiative transfer code presented by \citet{magee--18}, we calculated light curves and spectra for parameterised ejecta structures that were designed to broadly mimic predictions from theoretical explosion models (e.g. \citealt{kromer--10, polin--19, gronow--20}). We considered a range of white dwarf core masses (0.9 -- 1.2~\mass{}) and helium shell masses (0.01 -- 0.10~\mass{}), which effectively amounts to a range of \nickel{} masses. We also considered, for the first time, a large range of possible compositions for the burned material produced in the helium shell, which may result from different levels of pollution in the shell pre-explosion \citep{shen--09, waldman--11, kromer--10, gronow--20}. 

\par

Broadly, our model set may be separated into two categories: those that contain iron-group elements (IGE) in the shell and those that do not. Consistent with previous studies (e.g. \citealt{noebauer-17, jiang--2017, polin--19}), we find that those models containing IGE in the shell produce a bump in their respective light curves within the days following explosion. The luminosity and timescale of the bump can show considerable variation, reaching up to $M_{\rm{B}} \sim -18$ and lasting a few days for massive shells. Although the bump is most pronounced for bluer bands (e.g. $B$), it is also visible at longer wavelengths. At later times, light curves and spectra show extremely red colours and much of the flux below $\sim$4\,500~\AA\, has been suppressed. This also leads to fast declining light curves with rise times to $B$-band maximum typically around two weeks. Conversely, models that do not contain IGE in the shell show a relatively flat and blue colour evolution, and longer rise times that are more typical of normal SNe~Ia.

\par

As shown previously \citep{kromer--10, townsley--19}, models that do not contain IGE in the shell provide good agreement with observations of normal SNe~Ia around maximum light. Here, we have extended this and shown that the double detonation scenario is consistent with normal SNe~Ia beginning a few days after explosion. Our models do not provide evidence that the helium shell must contain specific elements (e.g. \sulphur{}), but rather show that it cannot contain IGE and beyond this requirement the composition has little effect. Therefore, provided the helium shell does not produce IGE during the explosion (which could be due to some amount of pollution), the double detonation scenario may be considered viable for a range of normal SNe~Ia and cannot be excluded.  Future explosion models should investigate this further by exploring a range of core and helium shell masses, as well as initial helium shell compositions.

\par

We also compared our models to SNe~Ia that show early bumps in their light curves. While we find that the bumps of all objects (with the exception of iPTF14atg) can be reproduced, only those objects with red colours at maximum light ($B-V \gtrsim 1$) are matched throughout their evolution. For blue objects, the model spectra at maximum light typically show broader features than observed, in addition to strong flux suppression. Regardless of the composition of the shell, we are unable to simultaneously match the early light curve and maximum light spectra of these blue events. The discovery of additional objects with early light curve bumps will help to determine the limit of the double-detonation scenario in reproducing observed light curve bumps. 

\par

Given that our double detonation models are unable to reproduce the complete evolution of blue SNe~Ia showing bumps at early times, this would indicate that an alternative source for the light curve bumps of these blue objects is necessary. Previous studies have also considered alternative scenarios and generally there is at least some disagreement between these scenarios and the observations. This may be due to either an over- or under-prediction of UV flux or the lack of features predicted by companion and CSM interaction scenarios in nebular spectra. It is therefore clear that there is much that remains unknown about the origin of the light curve bumps in SNe~Ia. As current and future facilities, such as the Zwicky Transient Facility and the Vera C. Rubin observatory (LSST), discover more SNe~Ia within hours of explosion, an investigation of general trends among the class will become possible. This should provide further insights into the nature of these enigmatic bumps.

\section*{Acknowledgements}

We thank the anonymous referee for their detailed and constructive comments, which helped to improve the clarity of our manuscript. We thank U. N\"obauer and A. Polin for providing densities and compositions for models used in their respective works. 
This work was supported by TCHPC (Research IT, Trinity College Dublin). Calculations were performed on the Kelvin cluster maintained by the Trinity Centre for High Performance Computing. This cluster was funded through grants from the Higher Education Authority, through its PRTLI program. This work made use of the Queen's University Belfast HPC Kelvin cluster. MM and KM are funded by the EU H2020 ERC grant no. 758638. This research made use of \textsc{Tardis}, a community-developed software package for spectral synthesis in supernovae
\citep{tardis}. The development of \textsc{Tardis} received support from the
Google Summer of Code initiative and from ESA's Summer of Code in Space program. \textsc{Tardis} makes extensive use of Astropy and PyNE.
This work made use of the Heidelberg Supernova Model Archive (HESMA), https://hesma.h-its.org.

\section*{Data Availability}

All models presented in this work are available on GitHub\footnote{\href{https://github.com/MarkMageeAstro/TURTLS-Light-curves}{https://github.com/MarkMageeAstro/TURTLS-Light-curves}}.



\bibliographystyle{mnras}
\bibliography{Magee}




\appendix
\section{Start time convergence test}
\label{sect:converge}
All models presented in this work were calculated assuming a start time of 0.5\,d after explosion, which is comparable to the decay timescale of some of the radioactive isotopes included in the model. For packets injected before the start of the simulation, diffusion relative to the matter in the ejecta is assumed to be negligible. The energy of these packets is also reduced to account for work done on the ejecta (see e.g. \citealt{lucy-05}).

\begin{figure}
\centering
\includegraphics[width=\columnwidth]{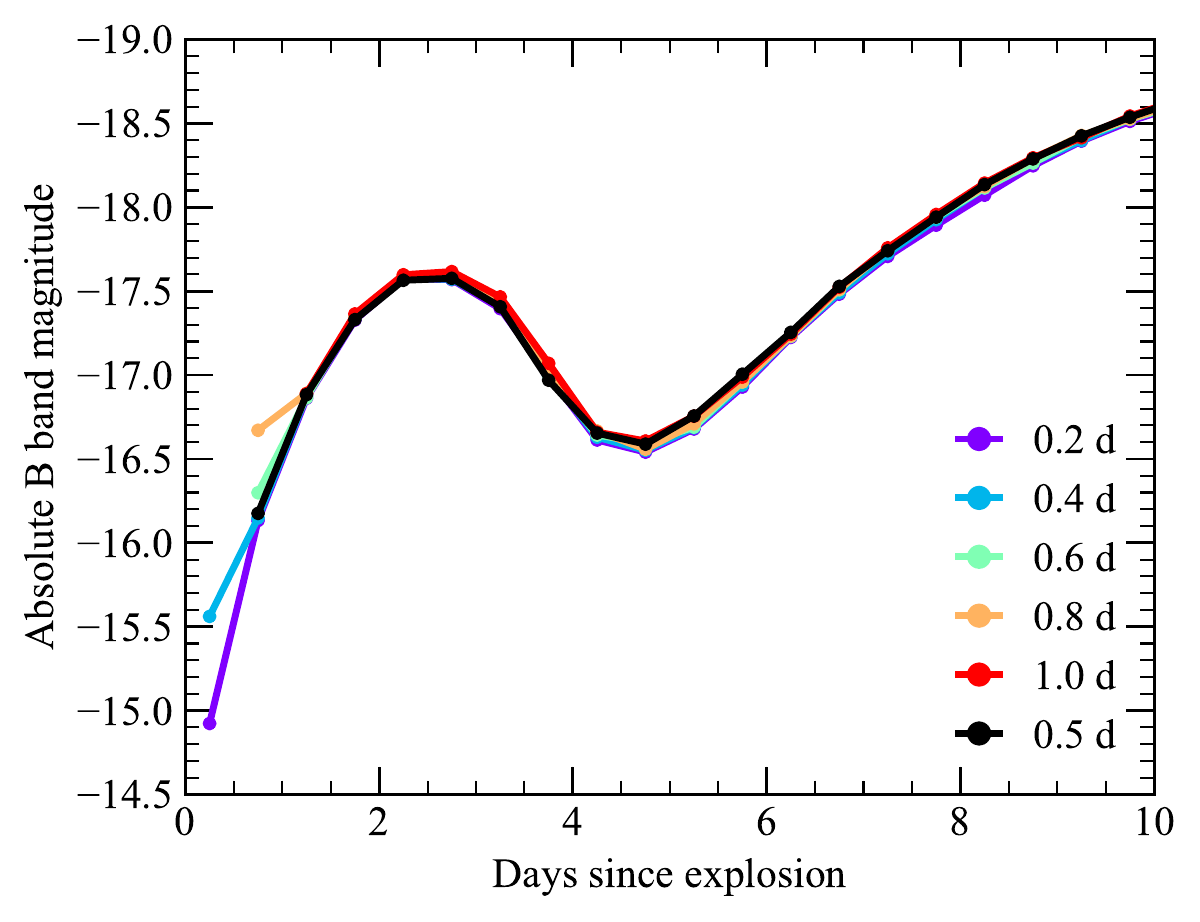}
\caption{Light curves calculated with different simulation start times for an \iron{}-dominated shell. Our nominal start time of 0.5\,d after explosion is shown in black.
}
\label{fig:converge}
\centering
\end{figure}

\par
In Fig.~\ref{fig:converge} we show \iron{}-dominated shell models calculated with earlier start times of 0.2\,d and 0.4\,d after explosion, as well as later times of 0.6\,d, 0.8\,d, and 1.0\,d after explosion. As demonstrated by Fig.~\ref{fig:converge}, the choice of 0.5\,d after explosion does not significantly impact the light curve. From $\sim$1\,d after explosion, all models produce comparable light curves to each other and show only minor variations ($\lesssim$0.05~mag.) consistent with Monte Carlo noise. Assuming a later start time of 0.8\,d after explosion does not reproduce the first light curve point of our nominal 0.5\,d case. For earlier start times however, the light curves show good agreement with our nominal case. Therefore, although the half-life of \iron{}, for example, is shorter than our 0.5\,d start time, the light curves in this case do not show significant variation when assuming earlier start times.

\section{Effects of core \nickel{} mass}
\label{sect:core_ni}

As discussed in Sect.~\ref{sect:construct_core_comp}, the \nickel{} masses of our models are based on those of similar models presented by \citet{kromer--10} and \citet{polin--19}. \citet{kushnir--20} present a detailed study of bare, sub-Chandrasekhar mass detonation models and show that the \nickel{} masses presented in these works may be systematically lower than those determined by other studies. Predicted \nickel{} masses calculated by \citet{kushnir--20} are shown in Fig.~\ref{fig:ni_masses}(a) for their default setup. Also shown are \nickel{} masses presented by \citet{shen--18} for white dwarfs in the range $\sim$0.9 -- 1.1~\mass{}, which show only minor variations for different metallicities and agree with those of \citet{kushnir--20}.

\par

\begin{figure}
\centering
\includegraphics[width=\columnwidth]{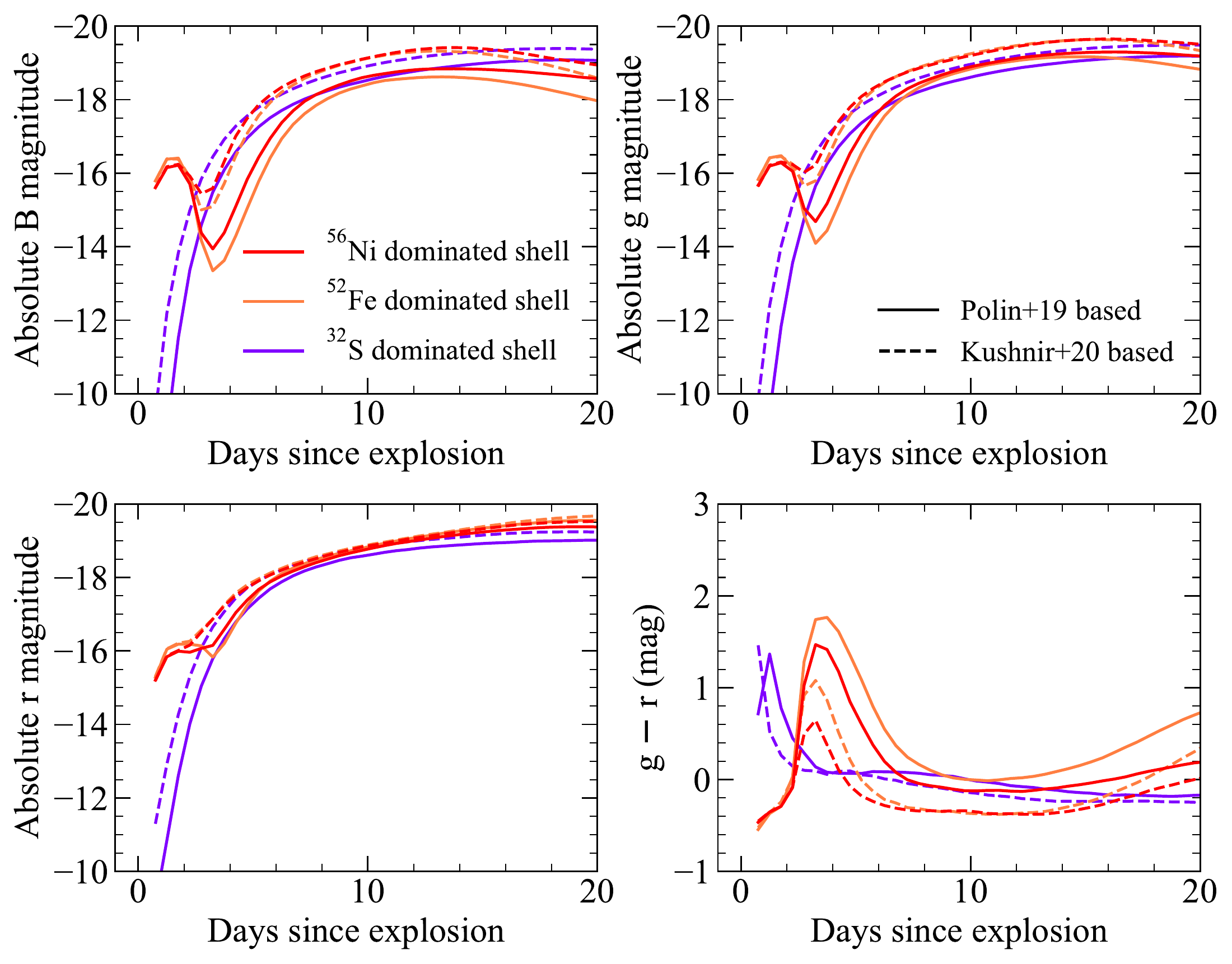}
\caption{Light curves and colours for models with different core \nickel{} masses. All models shown have a 1.0~\mass{} core and a 0.10~\mass{} shell, of which 50\% is burned to elements heavier than helium. We show models with dominant shell products of \iron{} and \sulphur{}, as representative of models with IGE- and IME-dominated shells. 
}
\label{fig:ni_mass_lc_dists}
\centering
\end{figure}

Figure~\ref{fig:ni_masses}(a) shows that differences in predicted \nickel{} masses may be relatively large -- particularly for lower mass white dwarfs. To account for this uncertainty in the \nickel{} mass produced, we present an additional set of models in which the core \nickel{} mass is based on the \citet{kushnir--20} models. Using a linear fit to these models, the core \nickel{} mass is given by:

\begin{equation}
\label{eqn:ni-mass-2}
    M(^{56}{\rm Ni}) = 2.6 \times (M_{\rm{core}} + M_{\rm{shell}} ) - 2.1,
\end{equation}
where $M_{\rm{core}}$ and $M_{\rm{shell}}$ are again the mass of the carbon-oxygen core and helium shell, respectively, in units of \mass{}. In Fig.~\ref{fig:ni_mass_lc_dists}, we show a comparison between models with \nickel{} masses based on \citet{polin--19} (Eqn.~\ref{eqn:ni-mass}) and those based on \citet{kushnir--20} (Eqn.~\ref{eqn:ni-mass-2}). As expected, those models with increased \nickel{} masses show systematically brighter peak luminosities, earlier rises, and overall bluer colours compared to their lower mass counterparts.

Figure~\ref{fig:ni_mass_lc_dists} shows that the increased \nickel{} mass results in brighter $B$- and $g$-band peaks by $\sim$0.35~mag for our \sulphur{}-dominated shell model. This model also begins to rise slightly earlier (by $\sim$0.5\,d), although there is not a significant shift in the time of peak brightness (i.e. $\lesssim$0.5\,d). For our \nickel{}- and \iron{}-dominated shell models, the changes from an increased \nickel{} mass are more dramatic, particularly in the $B$-band. In these cases, the $B$-band peaks are brighter by $\sim$0.6 -- 0.7~mag., while the $g$-band light curves experience slightly more modest increases of $\sim$0.4 -- 0.5~mag. Again, these models are brighter at early times, but there is no shift in the time of peak brightness. Although the shape of the main rising light curves differ due to the increased \nickel{} masses,  the shapes of the early light curve bumps are unaffected as these are primarily driven by the material in the shell. The rise time and magnitude of the bump peak is unaffected, however the decline after the peak of the bump is less pronounced. As the increased \nickel{} mass models begin to rise earlier, the difference between the peak of the bump and the minimum after the bump is reduced. This may make bumps less distinguishable in some cases as the earlier rise of the light curve produces more of a `shoulder' in the light curve than a well-defined rise and decline (such as those shown in the $r$-band in Fig.~\ref{fig:ni_mass_lc_dists}.

\par

\begin{figure}
\centering
\includegraphics[width=\columnwidth]{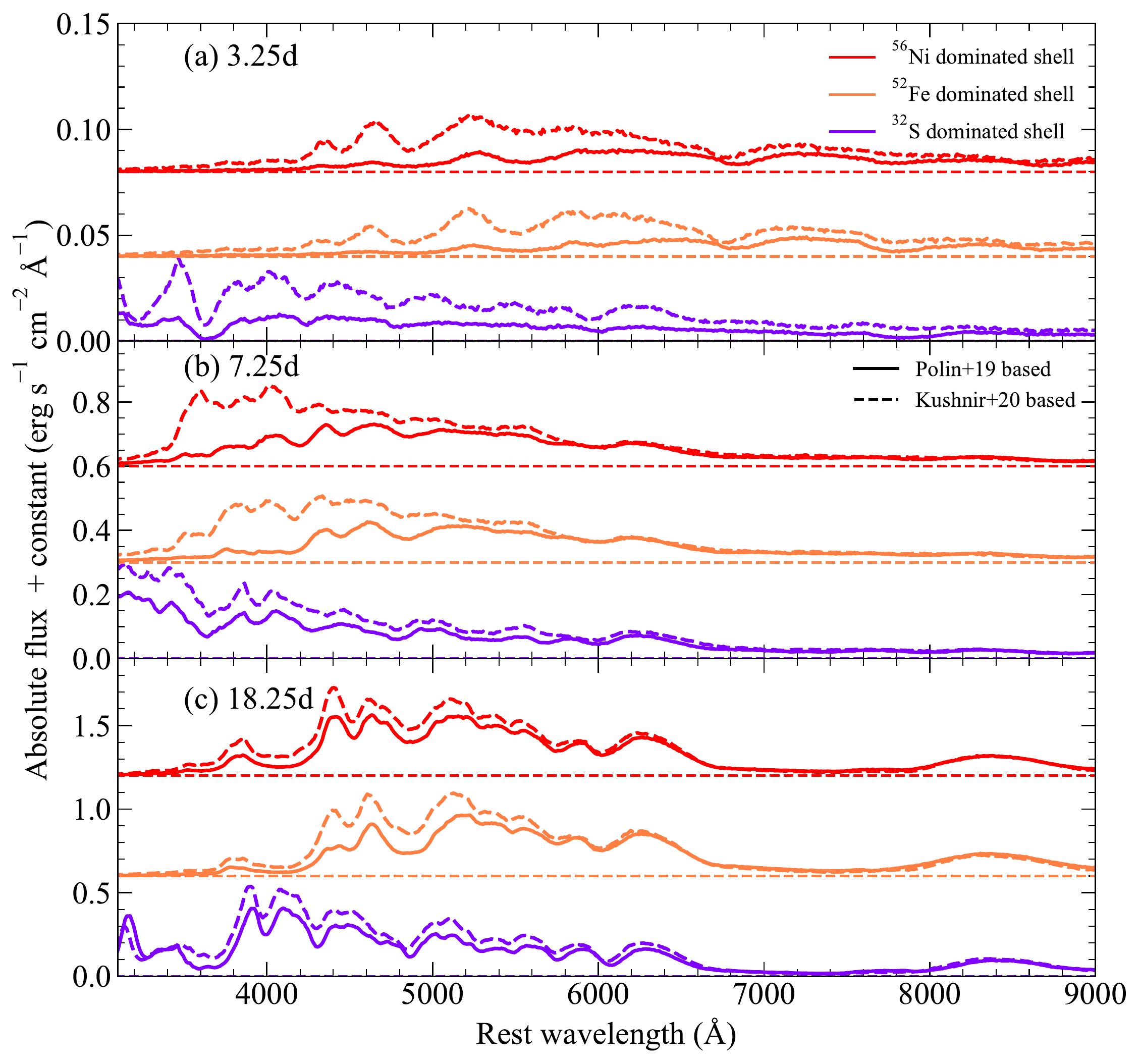}
\caption{Light curves and colours for models with different core \nickel{} masses. All models shown have a 1.0~\mass{} core and a 0.10~\mass{} shell, of which 50\% is burned to elements heavier than helium. We show models with dominant shell products of \iron{} and \sulphur{}, as representative of models with IGE- and IME-dominated shells. 
}
\label{fig:ni_mass_spec_dists}
\centering
\end{figure}

In Fig.~\ref{fig:ni_mass_lc_dists}, we also show the colour evolution for these models. This is further reflected in Fig.~\ref{fig:ni_mass_spec_dists}, which shows spectra for each model at 3.25\,d, 7.25\,d, and 18.25\,d after explosion. For our \sulphur{}-dominated shell model, the $g-r$ colour is bluer by $\lesssim$0.15~mag. throughout its evolution. This model also exhibits slightly higher velocities for the spectral features produced at all epochs. Again, the \nickel{}- and \iron{}-dominated shell models show larger changes. At their reddest points ($\sim$4\,d after explosion), the higher \nickel{} mass models show a shift to bluer $g-r$ colours between $\sim$0.6 -- 0.8~mag, while at their bluest points this shift is $\sim$0.3 -- 0.4~mag. Therefore the `red bump' (i.e. the transition a few days after explosion from blue colours to red, and back to blue again) because less pronounced and the $\Delta g-r$ for this transition decreases from $\sim$1.6 -- 1.7~mag. to $\sim$1.0 -- 1.4~mag. Interestingly, our \nickel{}- and \iron{}-dominated shell models with increased \nickel{} masses show a bluer $g-r$ colour than the \sulphur{}-dominated shell model between approximately one and two weeks after explosion. This is somewhat misleading, as Fig.~\ref{fig:ni_mass_spec_dists} shows that the spectra at these epochs for our \sulphur{}-dominated shell model are bluer and this is indeed reflected in the $U-B$ and $B-V$ colours. The appearance of the bluer $g-r$ colours for the \nickel{}- and \iron{}-dominated shells is likely due to increased fluorescence emission at these wavelengths. Similar to the \sulphur{}-dominated shell model, the increased \nickel{} mass also causes a slight shift to higher velocities for spectral features in the case of our \nickel{}- and \iron{}-dominated shell models.

\section{Effects of relative isotope abundances}
\label{sect:shell_distribution}

\begin{figure}
\centering
\includegraphics[width=\columnwidth]{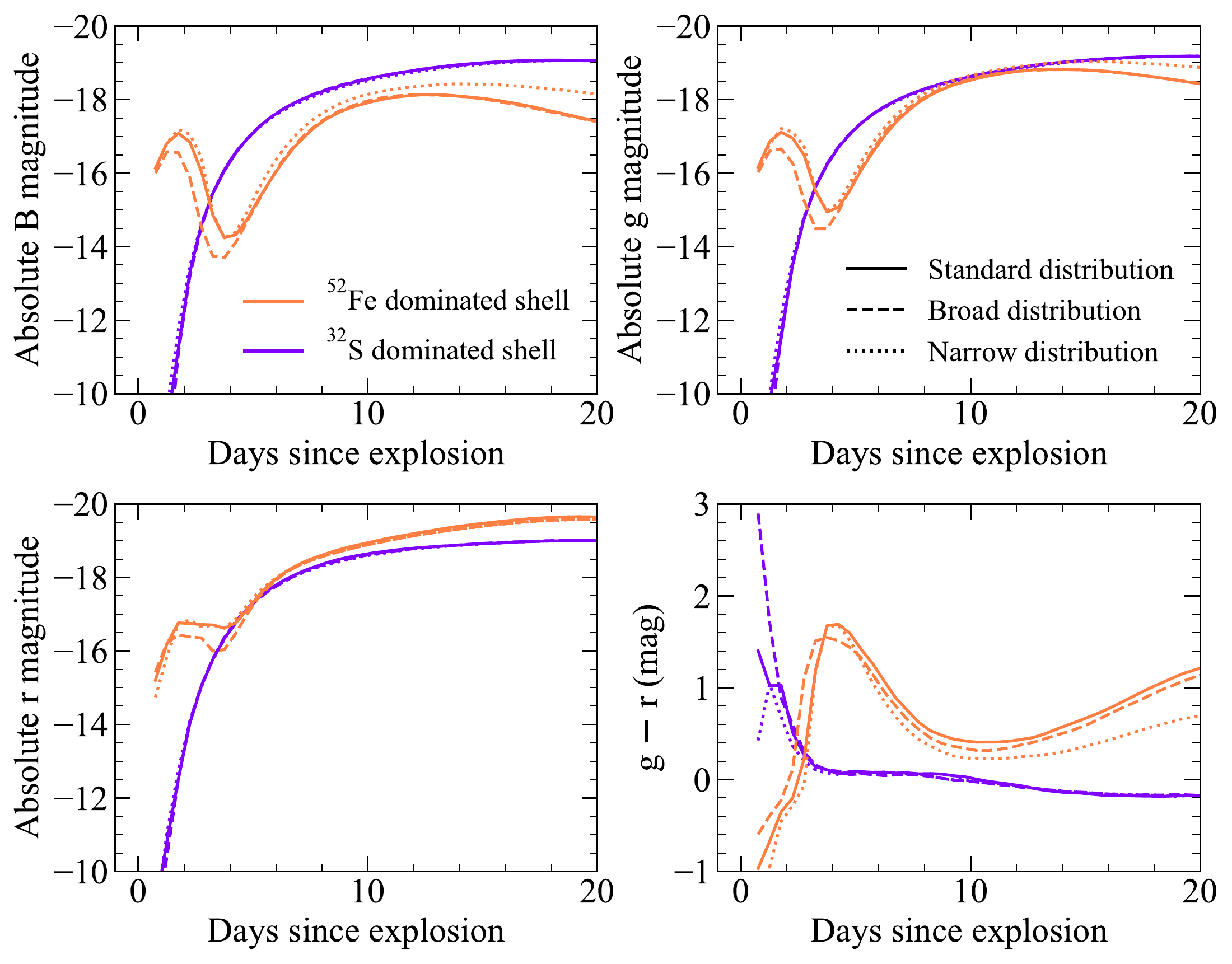}
\caption{Light curves and colours for models with different shell isotope distributions. All models shown have a 1.0~\mass{} core and a 0.04~\mass{} shell, of which 50\% is burned to elements heavier than helium. We show models with dominant shell products of \iron{} and \sulphur{}, as representative of models with IGE- and IME-dominated shells. 
}
\label{fig:shell_compositions_lc_dists}
\centering
\end{figure}

In Fig.~\ref{fig:shell_products}, we show the assumed relative abundances of isotopes in the helium shell for our models. In addition, we also show distributions in which the relative mass fraction of the dominant shell product decreases (broad distribution) or increases (narrow distribution). Here we discuss how these different distributions affect the light curves and spectra for our models.

\par

As demonstrated in Fig.~\ref{fig:shell_compositions_lc_dists}, changes in the relative abundances of isotopes in the helium shell have only minor effects on the light curves. For our \sulphur{}-dominated shell models, only a slight change in colour is observed at early times. For the \iron{}-dominated shell models, the effect is most pronounced in the $B$-band at early times. Relative to our standard case, the broad distribution shows a somewhat fainter early bump, which is not surprising given the decrease in mass fraction of radioactive isotopes. At later times, the narrow distribution is brighter than both our standard and broad distributions in the $B$-band. In the narrow distribution, as the mass fraction of \iron{} increases, the relative fractions of all other isotopes decrease. Therefore, that the narrow distribution is brighter at later times likely points to the decreased contribution to line blanketing from having fewer different elements present in the ejecta.

\begin{figure}
\centering
\includegraphics[width=\columnwidth]{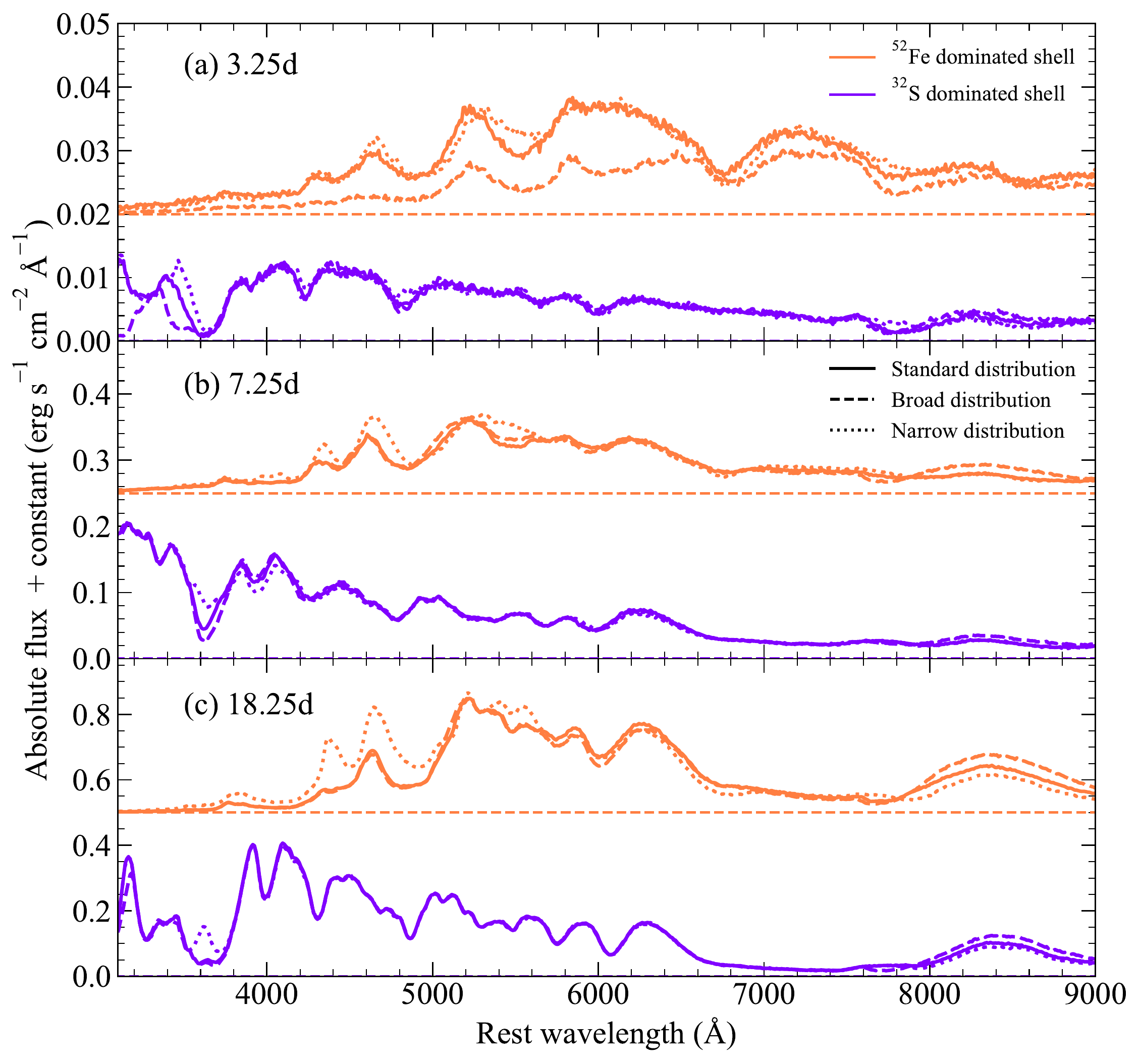}
\caption{Spectra for models with different shell isotope distributions. All models shown have a 1.0~\mass{} core and a 0.04~\mass{} shell, of which 50\% is burned to elements heavier than helium. We show models with dominant shell products of \iron{} and \sulphur{}, as representative of models with IGE- and IME-dominated shells. Spectra are shown at three epochs relative to explosion: 3.25\,d, 7.25\,d, and 18.25\,d.
}
\label{fig:shell_compositions_spectra_dists}
\centering
\end{figure}
\par

In Fig.~\ref{fig:shell_compositions_spectra_dists}, we show how the spectra are affected by changes in the relative abundances of the shell isotopes. Again, it is clear that our \sulphur{} dominated shell model shows only minor changes throughout its spectral evolution. At early times, the effect of a decreased \iron{} fraction is clearly apparent from the fainter and redder spectrum.


\bsp	
\label{lastpage}
\end{document}